\documentclass[a4paper]{article}
\usepackage[applemac]{inputenc}
\usepackage[T1]{fontenc}
\usepackage[english]{babel}
\usepackage{amssymb}
\usepackage{amsfonts}
\usepackage{amsmath}
\usepackage{amsthm}
\usepackage{subfigure}
\usepackage{xspace}
\usepackage{tabls}
\usepackage{graphicx}
\usepackage{bbold}
\usepackage{color}
\newcommand{\ignore}[1]{}

\newcommand{\wrt}{\emph{w.r.t.}\@\xspace}
\newcommand{\ie}{\emph{i.e.}\@\xspace}
\newcommand{\set}[1]{\left\{#1\right\}}
\newcommand{\card}[1]{\left|#1\right|}
\newcommand{\N}{\mathbb{N}}

\newcommand{\C}{\mathcal{C}}
\newcommand{\Cd}{\bigcup_{k\in\N}(\N^+)^k}
\newcommand{\og}[1]{\mathcal{G}_{#1}\xspace}
\newcommand{\fb}[1]{\bar{f}\left(#1\right)\xspace}
\newtheorem{theorem}{Theorem}
\newtheorem{corollary}[theorem]{Corollary}
\newtheorem{lemma}[theorem]{Lemma}
\newtheorem{proposition}[theorem]{Proposition}

\newtheorem{remark}{Remark}
\title{On symmetric sandpiles\thanks{A short version of this paper
has been presented at ACRI2006 conference.}} 
\author{Enrico Formenti
\footnote{Laboratoire I3S, Universit de Nice-Sophia Antipolis, Bt. ESSI, 930 route des Colles, 06903 Sophia Antipolis Cedex, \textsc{France}.\protect\\
\textsc{Email:}$\{$enrico.formenti,benoit.masson$\}$@unice.fr, pisokas@polytech.unice.fr}
 \and Benot Masson $^{\dag}$ \and Theophilos Pisokas $^{\dag}$}
\date{}

\begin{document}

\maketitle

\begin{abstract}
   A symmetric version of the well-known SPM model for sandpiles is introduced.
   We prove that the new model has fixed point dynamics. Although there might
   be several fixed points, a precise description of the fixed points is given. Moreover,
    we provide a simple closed formula for counting the number of fixed points 
    originated by initial conditions made of a single column of grains.
\end{abstract}

\section{Introduction}
Self-Organized Criticality (SOC) is a very common phenomenon which can be observed in Nature.
It concerns, for example, sandpiles formation, snow avalanches and so on~\cite{bak88}.

Practically speaking, it can be described as follows. Consider an evolving system. After a while, the
system reaches a \emph{critical state}. Any further move from this critical state will cause a deep
spontaneous reorganization of the whole system. No external parameter can be tuned to
control this reorganization. Thereafter, the system starts evolving to another critical state and
so on.

Sandpiles are a very useful model to illustrate SOC systems. Indeed, consider toppling sand grains
on a table, one by one. Little by little a sandpile will start growing and growing until the slope 
reaches a critical value. At this moment, any further addition of
a single sand grain will cause cascades of grains and deep reorganization of the whole pile.
Afterwards the sandpile restarts growing to another critical state and so on.

A formal model for sandpiles, called SPM, has been introduced in~\cite{goles93,goles02b,goles02}.
The sandpile is represented by a sequence of  ``columns''. Each column contains a certain number
of sand grains. The evolution is based on a local interaction rule (see Section~\ref{sec:spm}):
a sand grain falls from a column $A$ to its right neighbor $B$ if $A$ contains at least
two grains more than $B$; otherwise there is no movement.  The SPM model has been widely
studied~\cite{brylawski73,goles93,ruelle92,dhar95,moore99,miltersen99}. In particular, it 
has been proved that it has fixed point dynamics and a closed formula has been given to 
calculate precisely the length of the transient to the fixed point~\cite{goles93}.
Moreover, a precise description of the fixed point has been given~\cite{goles02b}.

All these results are very interesting but they have two main drawbacks.
First, they lack generality; indeed, the fixed point results are always obtained starting 
from very special initial sandpiles (just one column). In~\cite{formenti04,formenti05}, we tried to 
solve this problem by giving a fast algorithm for finding the fixed point starting from any possible initial
condition.
Second, the model lacks symmetry; in fact, grains either stay or move to the right only. Remark that
 in Nature, sandpiles evolve absolutely in a symmetrical manner. 
 
 In this paper we introduce SSPM: a symmetric version of SPM. The new model
 follows the rules of SPM but it applies them in both directions. For technical reasons
 that will be clearer later, we allow only one grain to move per time step.
 
 We prove that SSPM has fixed point dynamics. This is not a great surprise. To validate the
 new model, one should give a precise description of these fixed points and compare
 their ``shape'' with those of sandpiles in Nature.
 
 To this extent we use a formal construct which allows a better description of the dynamics: 
 orbit graphs. They are directed graphs of the relation ``being son of''. In Section~\ref{sec:og},
 the precise structure of their vertices  is given (under the condition of
 considering initial configurations made by a single column): a configuration belongs to
 some orbit graph if and only if it admits a crazed LR-decomposition 
 (see  Section~\ref{sec:og}).
 
 Practically speaking, a configuration admits a crazed LR-decomposition if it can be decomposed
 into an increasing part $L$ and a decreasing part $R$ and both in $L$ and in $R$ 
 any two plateaus (\ie consecutive columns of identical height) are separated by at least a ``cliff''
 (\ie consecutive columns with height difference strictly greater than $1$).
 
 The special structure of the vertices  allows a very useful description of the fixed points:
 they are configurations which admit a crazed LR-decomposition without cliffs.
 
 Finally, using this characterization of the ``shape'' of fixed points we provide a closed
 formula which computes the number of fixed points originated from the initial configuration
 $(n)$ (a single column containing $n$ grains). The surprise is that the formula
 is $\lfloor\sqrt{n}\rfloor$. Unfortunately, we have no practical or ``visual'' explanation for
 such a formula.
\section{The SPM model}\label{sec:spm}
A \emph{sandpile} is a finite sequence of integers $(c_1, \ldots,
c_k)$; $k\in\N$ is the \emph{length} of the pile.  Sometimes a
sandpile is also called a \emph{configuration}.  
Let $\C=\Cd$ be the set of all configurations.

Given a sandpile $(c_1, \ldots, c_k)$, the integer $n=\sum_{i=1}^k
c_i$ is the \emph{number of grains} of the pile.  Given a
configuration $(c_1, \ldots, c_k)$, a subsequence $c_i,\dots,c_j$
(with $1\leq i < j\leq k$) is a \emph{plateau} if $c_h=c_{h+1}$ for
$i\leq h<j$; $s=i-j+1$ is the length of the plateau and $p=c_i$ its
height.  A subsequence
$c_i,c_{i+1}$ is a \emph{cliff} if $c_i-c_{i+1}\geq 2$.

In the sequel, each sandpile $(c_1, \ldots, c_k)$ will be conveniently
represented on a two dimensional grid where $c_i$ is the grain content
of column $i$.

\begin{figure}[!ht]
  \begin{center}
   \includegraphics[scale=0.6]{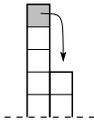}
  \end{center}
  \caption{local rule of SPM.}
  \label{fig:spm}
\end{figure}

\medskip
A \emph{sandpile system} is a finite set of rules that tell how the sandpile
is updated. SPM~\cite{goles93} (\emph{Sand Pile Model}) is the most known and the most simple sandpile system.
All initial configurations contain $n$ grains in the first column and nothing elsewhere \ie
they are of type $(n)$. It consists in only one local rule which moves a grain to the right whenever there is a cliff (see Figure~\ref{fig:spm}).

Formally, for any configuration $c$, if there exists $i \in \N$ such that $c_i-c_{i+1} \geq 2$, then $c$ evolves to $c'$ according the following relations:
\[
\left\{\begin{array}{rcl}
c'_i &=& c_i-1\\
c'_{i+1} &=& c_{i+1}+1\enspace.\\
\end{array}\right.
\]

This process is iterated until the rule cannot be applied anymore. We
say that a \emph{fixed point} is reached.

\begin{figure}[!ht]
  \begin{center}
  \includegraphics[width=\textwidth]{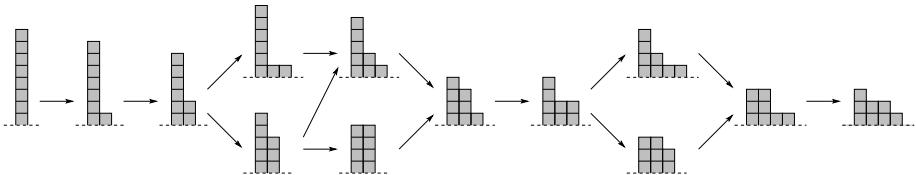}
  \end{center}
  \caption{$\og{(8)}$, orbit graph of a single pile with $8$ grains for SPM.}
  \label{fig:spm_lattice}
\end{figure}

Along the evolution of the pile, the rule may be applicable at
different places in the configuration. To illustrate this, we
represent the set of reachable configurations (starting from a single
column) on an oriented graph where the vertices are the
configurations. There is an edge between two configurations $c^1$ and
$c^2$ when $c^2$ can be obtained by applying the local rule somewhere
in $c^1$ (see Figure~\ref{fig:spm_lattice} for an example, starting
from a single column with 8 grains). This is called the \emph{orbit
  graph} of the initial configuration $c$, denoted by $\og{c}$.

The following theorem proves that the fixed point is unique,
independently of the order of application of the local rule.

\begin{theorem}[\cite{goles93}]\label{th:spm_lattice}
  For any integer $n$, $\og{(n)}$ for SPM is a lattice and is finite.
\end{theorem}

The following lemma characterizes the elements of the lattice.

\begin{lemma}[\cite{goles02b}] \label{lemm:cond1}
  Consider a configuration $c$ and let $n$ be its number of grains.
  Then, $c \in \og{(n)}$ for SPM if and only if it is decreasing and
  between any two plateaus of $c$ there is at least a cliff.
\end{lemma}

\begin{remark}
  Consider a configuration $c$, and assume that $c$ contains a plateau
  of length $3$. Such a plateau can be seen as two consecutive
  plateaus of length $2$. Thus, by Lemma~\ref{lemm:cond1}, $c$ does
  not belong to any orbit graph.
\end{remark}

From Lemma~\ref{lemm:cond1}, it is easy to see that a fixed point
$\Pi$ is a decreasing configuration with no cliffs and at most one
plateau. Therefore for any $n \in \N$, we can describe the fixed point
$\Pi$ of $(n)$ by

\[
\Pi = \left\{\begin{array}{l@{\hspace{1cm}}l}
      (p, p-1, \ldots, 1) & \text{if $q=0$}\enspace,\\
      (p, p-1, \ldots, q+1, q, q, q-1, \ldots, 1) & \text{otherwise,}
    \end{array}\right.
\]
where $\langle p,q \rangle$ is the unique decomposition of $n$ in its \emph{integer sum}:
\[
n = q + \sum_{i=1}^p i = q + \frac{p\cdot(p+1)}{2} \enspace.
\]

\section{The symmetric model}
%

\newcommand{\ap}{\ensuremath{c_{i+1}}}
\newcommand{\am}{\ensuremath{c_{i-1}}}
\newcommand{\dl}{\ensuremath{\delta^l}}
\newcommand{\dr}{\ensuremath{\delta^r}}
\newcommand{\Vp}{\ensuremath{V^r}}
\newcommand{\Vm}{\ensuremath{V^l}}
\newcommand{\Vpm}{\ensuremath{V^{lr}}}
\newcommand{\eps}{\ensuremath{\varepsilon}}
\newcommand{\ohm}{\ensuremath{\omega}}
\newcommand{\ra}{\ensuremath{\rightarrow}}
%

In this section we extend SPM to SSPM (Symmetric SPM) according to the
following guidelines:
\begin{itemize}
\item a grain can move either to the left or to the right, if the
  difference is more than $2$;
\item when a grain can move only in one direction, it follows the SPM
  rule (right) or its symmetric (left).
\end{itemize}

For all configurations $c=(c_1, \ldots, c_k)$,
the following local rules formalize the above requirements:
\[
\Vp_i(c_1, \ldots, c_k)=
\begin{cases}
  (c_1, \ldots, c_i\mathbf{-1}, \ap\mathbf{+1}, \ldots,c_k)&\text{if}\;i\ne k\enspace,\\
  ( c_1, \ldots, c_k\mathbf{-1}, \mathbf{1})&\text{otherwise},\\  
\end{cases} 
\]
\[ 
  \Vm_i(c_1, \ldots, c_k)=
\begin{cases}
  (c_1, \ldots, \am\mathbf{+1}, c_i\mathbf{-1}, \ldots, c_k)&\text{if}\;i\ne 1\enspace,\\
  (\mathbf{1},  c_1\mathbf{-1}, \ldots, c_k)&\text{otherwise.}\\
\end{cases}  
\]

Let $\dr_i(c)$ denote the \emph{difference} between the grain content
of column $i$ and the one of column $i+1$ of $c$; define
$\dr_k(c)=c_k$. Similarly, $\dl_i(c)$ denotes the \emph{difference}
between the grain content of column $i$ and the one of column $i-1$ of
$c$ with $\dl_1(c)=c_1$.

\paragraph{Notation.} For $a,b\in\N$ with $a<b$, let $[a,b]$ denote
the set of integers between $a$ and $b$.  \smallskip

From the local rule we can define a \emph{next step rule}
$\bar{f}:\C\mapsto\mathfrak{P}(\C)$ as follows
\[
\fb{c} = \big\{ V^r_i(c) \;|\; \delta^r_i (c)\geq 2, i \in[1,k]\big\}
\cup \big\{ V^l_i(c) \;|\; \delta^l_i(c) \geq 2, i \in[1,k]\big\}\enspace.
\]

Finally, using the next step rule, one can define the \emph{global
  rule} which describes the evolution of the system from time step $t$
to time step $t+1$ :
\[
\forall S\in\mathfrak{P}(\C),\;f(S)=\bigcup_{c\in S}\fb{c}\enspace.
\]

When no local rule is applicable to $c$, \ie $f(\{c\})=\emptyset$, we
say that $c$ is a \emph{fixed point of SSPM}. For $n\in\N$, let
$f^n$ denote the $n$-th composition of $f$ with itself.
\smallskip

The notion of orbit graph can be naturally extended to the symmetric
case by using the functions $\Vp_i$ and $\Vm_i$.
In the sequel, when speaking of orbit graph, we will always mean
the orbit graph \wrt the SSPM model.
\subsection{Fixed point dynamics}

In this section we prove that SSPM has fixed points dynamics. This
result is obtained by using a ``potential energy function'' and by
showing that this function is positive and non-increasing.
\smallskip

Given a configuration $c=(c_1, \ldots, c_k)$, the \emph{energy} of a column 
$c_i$ ($i\in[1,k]$) is defined as follows
\[
	\eps(c_i) = \sum_{j=1}^{c_i} j\enspace.
\]
Therefore, the \emph{total energy} of a configuration $c=(c_1, \ldots, c_k)$ is naturally 
defined as 
\[
	E(c) = \sum_{i=1}^k \eps(c_i)\enspace.
\]

\begin{lemma}\label{lem:pre.nrg}
  Consider a configuration $c=(c_1, \ldots, c_k)$ with $n$ grains.
  Then it holds that $E(c)\leq E((n))$; equality holds if and only if
  $c=(n)$.
\end{lemma}
\begin{proof}
   Remark that $n=\sum_{i=1}^k c_i$.  
   Define $h(i)=\sum_{j=c_1+c_2+\cdots+c_{i-1}+1}^{c_1+c_2+\cdots+c_i} j$.
   Then, $E((n))=\sum_{j=1}^n j$ can be rewritten as
   $E((n))=\sum_{i=1}^k h(i)$. Note that $h(i)\geq\epsilon(c_i)$ for any
   $i\in[1,k]$; equality holds if and only if $i=1$.
\end{proof}

The function $E$ can be naturally extended to work on set of configurations as follows
\[
\forall S\in\mathfrak{P}(\C),\;E(S)=\max\set{E(c),\,c\in S}\enspace,
\]
with $E(\emptyset)=0$.

The following lemma is straightforward from the definition of the energy function.
\begin{lemma}\label{lem:Ef.inf.E}
  For any set of configurations $S\ne\emptyset$, $E(f(S))<E(S)$.
\end{lemma}

The following simple proposition describes the general structure of
the orbit graph.

\begin{proposition}\label{prop:fixp}
  For any initial configuration $c$, $\og{c}$ is finite, contains at
  least a fixed point but no cycles.
\end{proposition}
\begin{proof}
  If $\fb{c}=\emptyset$ \ie c is a fixed point, then we are done.
  Assume that $c$ is not a fixed point. Remark that its energy is
  finite. By Lemma~\ref{lem:Ef.inf.E}, it holds that
  $E(\set{c})>E(f(\set{c}))$ and $E(f^t(\set{c}))>E(f^{t+1}(\set{c}))$
  for $t>1$ (unless $f^t(\set{c})=\emptyset$).  Since $E$ is a
  positive function, there must exist $h\in\N$ such that
  $f^h(\set{c})=\emptyset$.  Then, $f^{h-1}(\set{c})$ contains a fixed
  point. If $f^h(\set{c})=\emptyset$, then the orbit graph $\og{c}$ is
  finite, since $|f^t(\set{c})|$ is finite for any $t\in\N$.
  
  Finally, there are no cycles in $\og{c}$ otherwise the elements of
  the cycle would contradict Lemma~\ref{lem:Ef.inf.E}.
\end{proof}

The following corollary is given only to further stress the result
of Proposition~\ref{prop:fixp}.

\begin{corollary}\label{cor:fixpstress}
   SSPM has fixed point dynamics.
\end{corollary}

\begin{figure}[!ht]
  \begin{center}
    \includegraphics[scale=0.3]{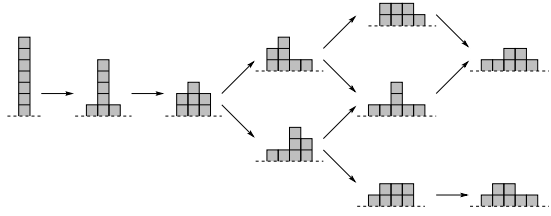}
  \end{center}
  \caption{$\og{(5)}$, orbit graph of a single pile with $5$ grains for SSPM. Remark that
  there are two distinct fixed points.}
  \label{fig:sspm.og5}
\end{figure}

Corollary~\ref{cor:fixpstress} says that independently of the order
of application of local rules both with respect to type of rule and to
the application site, SSPM evolves towards a fixed point. The problem
is that this fixed point might not be unique.
Figure~\ref{fig:sspm.og5} gives an example of this fact.

 Despite the non-uniqueness, in the next section we give a precise characterization
 of the structure of the fixed points. This characterization is essentially deduced from
 the properties of the vertices of the orbit graphs.
\subsection{Orbit graphs}\label{sec:og}
In~\cite{goles02b}, the authors precisely described the structure of the orbit graph
of SPM when started on initial condition $(n)$. They proved that it is the graph of
a lattice. As a consequence, they deduced the uniqueness of the fixed point for SPM.

We have already seen that in the SSPM case, the dynamics is of fixed point type,
but the fixed point might not be unique. Hence, it is clear that the orbit graph of SSPM
 is no more the graph of a lattice. In this section, we detail the overall structure of the
vertices of these graphs.
\smallskip

A configuration $c=(c_1,c_2,\ldots,c_k)$ is \emph{LR-decomposable} if it  
can be divided into two \emph{zones}: $L(c)=[1,t],  R(c)=[t+1,k]$ such that
\begin{enumerate}
   \item $\forall i\in L(c), i \ne t, c_i\leq c_{i+1}$ \ie $L(c)$ is non-decreasing;
   \item $\forall i\in R(c), i \ne k, c_{i}\geq c_{i+1}$ \ie $R(c)$ is non-increasing.
\end{enumerate}

\begin{figure}[!ht]
  \begin{center}
    \subfigure[LR-decomposition.]{
      \includegraphics[width=0.25\textwidth]{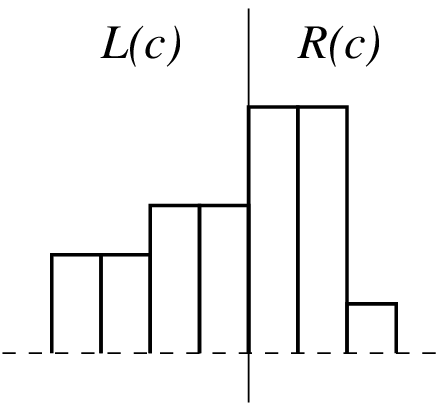}
      \label{fig:LR}
    }\hspace{5mm}
    \subfigure[The top of $c$.]{
      \includegraphics[width=0.25\textwidth]{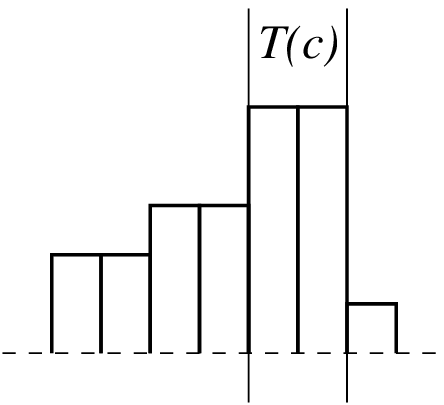}
      \label{fig:T}
    }
    \caption{decomposition of a configuration $c$.}
    \label{fig:decomp}
  \end{center}
\end{figure}

Figure~\ref{fig:LR} give an example of LR-decomposition.
For any configuration $c$, let $T(c)=\set{i\in[1,k], \forall j\in[1,k], c_i\geq c_j}$.
In the sequel, $T(c)$ is called the \emph{top} of $c$, see Figure~\ref{fig:T}.

Given a configuration $c=(c_1,c_2,\ldots,c_k)$, a set of 
consecutive indexes
$I\subseteq[1,k]$ is \emph{crazed} if any two plateaus in $I$ are
separated by at least a cliff.  A configuration $c$ has a
\emph{crazed} LR-decomposition if it admits a LR-decomposition in
which both $R(c)$ and $L(c)$ are crazed.
\smallskip

A configuration might have several different LR-decompositions.  The
following propositions tell which of them we are interested in. The
proof of  Proposition~\ref{prop:crazedLTR} will be made progressively, using
several technical lemmas.

\begin{lemma}\label{lem:LTR}
 Consider  $n\in\N$ and $c\in\og{(n)}$. Then $c$ is LR-decomposable. 
\end{lemma}
\begin{proof}
  The thesis is trivially true for the initial configuration $(n)$.
  Now assume that $c\in\og{(n)}$.  Let $L(c)=[1,t], R(c)=[t+1,k]$ be a
  LR-decomposition of $c$.  Consider $d\in\fb{c}$ and assume that
  $d=\Vp_i(c)$. We have three cases:
  \begin{enumerate}
  \item $i=t$; then $L(d)=L(c)\setminus\set{t}$ and
    $R(d)=R(c)\cup\set{t}$.  $L(d)$ is non-decreasing since
    $L(d)\subseteq L(c)$. $R(d)$ is non-increasing since $d_t\geq
    d_{t+1}$ and $R(c)\subseteq R(d)$;
  \item $i\in[t+1,k-1]$; then $L(d)=L(c)$ and $R(d)=R(c)$. Of course
    $L(d)$ is non-decreasing and $R(d)$ is non-increasing.
  \item $i=k$; then $L(d)=L(c)$ and $R(d)=R(c)\cup\set{k+1}$. Of
    course $L(d)$ is non-decreasing. $R(d)$ is non-increasing since
    $R(c)\subseteq R(d)$ and $d_k\geq 1$.
  \end{enumerate}
  The proof is similar if $d=\Vm_i(c)$.
  \end{proof}

\begin{lemma}\label{lem:pl3}
  Consider $n\in\N$ and $c\in\og{(n)}$. Let $T(c)$ be the top of $c$.
  Any LR-decomposition of $c$ is such that both $L(c)\setminus T(c)$
  and $R(c)\setminus T(c)$ have no plateaus of size strictly greater
  than $2$.
\end{lemma}
\begin{proof}
  Consider $n\in\N$ and $c\in\og{(n)}$.  Remark that if $c=(n)$ then
  the thesis is true. Now, assume that $c\ne(n)$, then $c$ should have
  an ancestor in $\og{(n)}$.  We prove the thesis for $R(c)\setminus
  T(c)$ by contradiction. Assume that there exist a plateau of size
  $m>2$ in $R(c)\setminus T(c)$ \ie there exists $i\in R(c)\setminus
  T(c)$ such that $c_i=c_{i+1}=\ldots=c_{i+m-1}$. By the hypothesis we
  know that $c_{i-1}>c_i$ and $c_{i+m-1}>c_m$ (we assume $c_{k+1}=0$).
   
  Consider a configuration $d$ such that $c\in\fb{d}$ and $c=\Vp_j(d)$
  for $j\in [i-1,i+m-1]$.  Then, $d$ is not LR-decomposable and, by
  Lemma~\ref{lem:LTR}, it does not belong to $\og{(n)}$. A
  configuration $d$ such that $c=\Vm_j(d)$ for $j\in [i,i+m]$ is not
  LR-decomposable either, $d \not\in \og{(n)}$. The proof for
  $L(c)\setminus T(c)$ is very similar.
\end{proof}

\begin{lemma} \label{lem:pl2}
 Consider $n\in\N$ and $c\in\og{(n)}$.  Let $T(c)$ be the top of $c$.
   Any LR-decomposition of $c$ is such that both $L(c)\setminus T(c)$
   and $R(c)\setminus T(c)$ are crazed.
\end{lemma}
\begin{proof}
  Let $c$ be a configuration where $R(c) \setminus T(c)$ is not
  crazed, \ie $c$ contains two plateaus not separated by a cliff. Then
  there are two indices $i, j \in [t, k]$, $i < j$, such that $c_{i-1}
  > c_i=c_{i+1}$, $c_j=c_{j+1} > c_{j+2}$ and for all $h \in [i+1,
  j-1]$, $c_h = c_{h+1}+1$. Let $\delta = j-i$, we prove that $c
  \not\in\og{(n)}$ by induction on $\delta$. If $\delta=1$,
  Lemma~\ref{lem:pl3} proves the thesis. Suppose the result is true
  for every $\delta \in [1,m]$, and that we have $\delta = m+1$. Let
  $c$ be a configuration which contains two plateaus not separated by
  a cliff, at distance $j-i = m+1$. Consider a configuration $d$ such
  that $c \in \fb{d}$ and $c = V^r_h(d)$ for $h \in [i-1,j+1]$. We
  have the following cases:
  \begin{itemize}
  \item $h=i-1$; then $d_{i-1} > d_i < d_{i+1}$, $d$ is not
    LR-decomposable, $d \not\in \og{(n)}$ by Lemma~\ref{lem:LTR};
  \item $h\in[i,j-2]$; then there is a plateau in $d$ at position $h+1$ and
    the plateau at position $j$ is unchanged, by induction over
    $\delta$ it holds that $d \not\in \og{(n)}$;
  \item $h=j-1$; $d_{j-1} > d_j < d_{j+1}$, $d$ is not
    LR-decomposable, Lemma~\ref{lem:LTR} says that $d \not\in \og{(n)}$;
  \item $h=j$; there are two plateaus in $d$ at position $j-1$ and
    $i$, we have that $d \not\in \og{(n)}$ by induction over $\delta$;
  \item $h=j+1$; $d_{j-1} > d_j < d_{j+1}$, $d$ is not
    LR-decomposable, from Lemma~\ref{lem:LTR} $d \not\in \og{(n)}$.
  \end{itemize}
  Choose $d$ such that $c \in \fb{d}$ and $c = V^l_h(d)$ for $h \in
  [i,j+2]$. The only possible values for $h$ are the non-decreasing
  parts in $c$, \ie $h \in \{i+1, j+1\}$:
  \begin{itemize}
  \item if $h=i+1$; there are two plateaus in $d$ at position $i+1$
    and j, by induction over $\delta$ it holds that $d \not\in
    \og{(n)}$;
  \item if $h=j+1$; $d_{j-1} > d_j < d_{j+1}$ and from Lemma~\ref{lem:LTR},
    $d \not\in \og{(n)}$.
  \end{itemize}
  Therefore among all the ancestors of $c$ which create the plateaus,
  none is in $\og{(n)}$, hence $c \not\in \og{(n)}$. A similar proof
  can be done if $L(c) \setminus T(c)$ is not crazed.
\end{proof}

It is obvious that the cardinality of $T(c)$ is bigger of equal to $1$
for all configurations.  Using very simple examples one can verify
that $|T(c)|$ can also be equal to $2$, $3$ or $4$. The following
result proves that these are the only possible values for the
cardinality of $T(c)$ when $c$ belongs to an orbit graph.

\begin{lemma} \label{lem:T4}
   Consider  $n\in\N$ and $c\in\og{(n)}$. Then $|T(c)|\leq 4$.
\end{lemma}
\begin{proof}
  If $c=(n)$ then the thesis trivially holds. Assume that $c\ne(n)$, then
   $c$ should have an ancestor in $\og{(n)}$. By contradiction,
  let $\card{T(c)}> 4$.
    
  Consider a LR-decomposable configuration $d$ such that
  $c\in\Vm_i(d)$ for $i\in T(c)$.  Then, according to
  Lemma~\ref{lem:LTR}, $d$ has only two possible pre-images
   \begin{align*}
      1)&\;\ldots,d_l-1,d_l+1,d_{l+1},d_{l+2},d_{l+3},\ldots&\text{with}&\;
      d_{l+1}=d_{l+2}=d_{l+3}\enspace;\\
      2)&\;\ldots,d_{m-4},d_{m-3},d_{m-2},d_{m-1}+1,d_{m}-1,\ldots&\text{with}&\;
      d_{m-4}=d_{m-3}=d_{m-2}\enspace;
   \end{align*}
   where $l=\min T(d)$ and $m=\max T(d)$. Both $1)$ and $2)$ contradict
   Lemma~\ref{lem:pl3}.
\end{proof}

The following proposition gives a precise characterization of the
configurations of the orbit graph.  Its proof is very technical as
many different cases have to be considered, but each of them is solved
quite simply using the previous lemmas.

\begin{proposition} \label{prop:crazedLTR}
  Consider $n\in\N$ and $c\in\og{(n)}$. Then $c$ has a crazed
  LR-de\-com\-po\-si\-tion.
\end{proposition}
\begin{proof}
  Let $n\in\N$, $c=(c_1, \ldots, c_k)\in\og{(n)}$ and $T(c)$ be the
  top of $c$. There are 4 cases, depending on the cardinality of
  $T(c)$ (Lemma~\ref{lem:T4}).

  \begin{itemize}
  \item If $|T(c)| = 1$, then by Lemma~\ref{lem:pl2} any
    LR-decomposition of $c$ into $L(c)$ and $R(c)$ is crazed (there
    are no additional plateaus in $T(c)$).

  \item If $|T(c)| = 2$, choose $L(c) = [1,t]$ and $R(c) = [t+1, k]$
    where $T(c) = [t, t+1]$. Since $c$ is LR-decomposable
    (Lemma~\ref{lem:LTR}), this is a valid LR-decomposition.  Again
    there are no plateaus in $L(c) \cap T(c)$ and in $R(c) \cap T(c)$,
    hence this is a crazed LR-decomposition (Lemma~\ref{lem:pl2}).

  \item If $|T(c)| = 4$, let $t \in [1,k]$ such that $T(c) = \{t, t+1,
    t+2, t+3\}$. Choose $L(c)=[1,t+1]$ and $R(c) = [t+2,k]$. Again, it
    is clearly a valid LR-decomposition. We prove that both $L(c)$ and
    $R(c)$ are crazed. The plateau $T(c)$ can be obtained from the
    configurations $d \in \og{(n)}$ such that $c=V^r_i(d)$, $i \in [t,
    t+3]$:
    \begin{itemize}
    \item if $i=t$ or $i=t+1$; then $d_i > d_{i+1} < d_{i+2}$, which
      is not possible because of Lemma~\ref{lem:LTR};
    \item if $i=t+3$; $d_t = d_{t+1} = d_{t+2} < d_{t+3}$, which is
      not possible because of Lemma~\ref{lem:pl3}.
    \item if $i=t+2$; $d_t = d_{t+1} < d_{t+2}$ and because of
      Lemma~\ref{lem:pl2}, $L(d) = [1, t+2]$ is crazed since $L(d)
      \cap T(d) = \emptyset$. Therefore $L(c)$ is crazed as
      $L(c)=L(d)$ and for all $i \in L(c)$, $c_i=d_i$. For $R(c)$,
      there are two cases.
      \begin{itemize}
      \item Either $c_{t+3} \geq c_{t+4}+2$; then the plateau
        $c_{t+2}, c_{t+3}$ in $R(c) \cap T(c)$ is separated from any
        other plateau in $R(c)$ by the cliff at position $t+3$, hence
        $R(c)$ is crazed.
      \item Or $c_{t+3} = c_{t+4}+1$; then $d_{t+2} > d_{t+3} =
        d_{t+4}$ and $[t+3, k]$ is crazed in $d$. This means that if
        $j \geq t+4$ is the lowest index such that $d_j = d_{j+1}$,
        there is a cliff somewhere at index $h$, $t+4 \leq h < j$
        (Lemma~\ref{lem:pl2}).  Hence this cliff is also in $c$, and
        there is no plateau in $c$ between $t+3$ and $h$. Therefore
        the plateau $c_{t+2}, c_{t+3}$ is separated from any other
        plateau in $R(c)$ by the cliff at index $h$, $R(c)$ is crazed.
      \end{itemize}
    \end{itemize}
    Similar results hold if $c=V^l_i(d)$, $i \in [t, t+3]$. Therefore
    there are only two possibilities for $d$, and for both of them
    $L(c)$ and $R(c)$ are crazed.

  \item If $|T(c)| = 3$, let $T(c) = \{t, t+1, t+2\}$. Suppose that
    $X(c) = [t+1, k]$ is crazed, then let $L(c) = [1, t]$ and $R(c) =
    X(c)$. This is a LR-decomposition of $c$ (Lemma~\ref{lem:LTR}), it
    is crazed because $L(c) \cap T(c)$ is also crazed
    (Lemma~\ref{lem:pl2}).

    If $X(c)$ is not crazed, clearly $[t+2, k]$ is crazed because of
    Lemma~\ref{lem:pl2}. We need to prove that $Y(c) = [1, t+1]$ is
    necessarily crazed. Let $j \in [t+2,k]$ be the lowest index such
    that $c_j = c_{j+1}$. Remark that for all $h \in [t+2,j-1]$,
    $c_h=c_{h+1}+1$. If $j=t+2$, we are in the case $|T(c)|=4$, solved
    previously. Otherwise, for any ancestor $d$ of $c$ such that
    $c=V^r_h(d)$, $h \in [t,j+1]$, it holds that
    \begin{itemize}
    \item if $h=t$; $d_{t} > d_{t+1} < d_{t+2}$ which is impossible
      because of Lemma~\ref{lem:LTR};
    \item if $h \in [t+1, j-1]$; $d_{h} > d_{h+1} = d_{h+2}$, from
      Lemma~\ref{lem:pl2} this is not possible;
    \item if $h=j$, $d_{j-1} > d_{j} < d_{j+1}$, Lemma~\ref{lem:LTR}
      proves that it is impossible;
    \item if $h=j+1$, plateau at $d_{j-1}, d_j$, by induction (see
      proof of Lemma~\ref{lem:pl2}) it leads to the case $|T(e)|=4$
      for an ancestor $e$ of $c$, which implies that $Y(c)$ is crazed.
    \end{itemize}
    If $d$ is such that $c=V^l_h(d)$, $h \in \{t, t+1, t+2, j+1\}$, it
    holds that
    \begin{itemize}
    \item if $h=t$; $d_t > d_{t+1} = d_{t+2}$, by Lemma~\ref{lem:pl2}
      this is impossible;
    \item if $h=t+1$; the proof is exactly the same as for the sub-case
      $i=t+2$ of the case $|T(C)| = 4$: $Y(c)$ is crazed;
    \item if $h=t+2$ or $h=j+1$; $d_{h-2} > d_{h-1} < d_{h}$ which is
      impossible (Lemma~\ref{lem:LTR}).
    \end{itemize}
    Therefore $Y(c), X(c)$ is a valid crazed LR-decomposition of $c$.
  \end{itemize}
\end{proof}

The converse of Proposition~\ref{prop:crazedLTR} is proved using 
another technical lemma.

\begin{lemma}\label{lem:cracra-1}
  Consider a configuration $c$, $c \ne (n)$ for all $n \in \N$, which
  admits a crazed LR-decomposition.  Then, there exists $d$ such that
  $c\in f(\set{d})$ and $d$ admits a crazed LR-decomposition.
\end{lemma}
\begin{proof}
  Assume that $c=(c_1, \ldots, c_k)$ is such that $c\ne(n)$ for
  $n=\sum_{i=1}^k c_i$. Moreover, assume that $c$ admits a crazed
  LR-decomposition and denote it by $L(c)$ and $R(c)$.

  If $L(c)=\emptyset$, then if $|R(c)| = 1$, nothing can be done: this
  is the case $c=(n)$, which is not possible by hypothesis. Otherwise,
  build a configuration $d$ as follows: $d_1=c_1+1$, $d_2=c_2-1$ and
  $d_i=c_i$ for $i\in[3,k]$. Then, $c=\Vp_1(d)$ and $L(d)=\emptyset$,
  $R(d)=R(c)$ is a crazed LR-decomposition of $d$. Note that if $c_2 =
  1$, $d_2 = 0$ so $d$ has length $k-1$. In that particular case, we
  define $L(d)=\emptyset$, $R(d)=[1,k-1]$.

  If $|L(c) = 1|$, we have two cases. If $c_1 > c_2$, we could have
  chosen $L(c) = \emptyset$ and $R(c) = [1, k]$, case solved
  previously. Otherwise, define $d$ such that $d_1 = c_1-1$,
  $d_2=c_2+1$ and $d_i=c_{i}$ for $i\in[3,k]$. It holds that
  $c=\Vm_2(d)$, and $L(d)=L(c)$, $R(d)=R(c)$ is a crazed
  LR-decomposition of $d$. Remark that if $c_1 = 1$, $d_1 = 0$ so $d$
  has to be shifted by $1$ to the left. In that particular case also,
  we define $L(d)=\emptyset$, $R(d)=[1,k-1]$ which is a crazed
  LR-decomposition of $d$.

  If $|L(c)|>1$, then we have two cases.
  \begin{itemize}
  \item $L(c)$ contains at least one plateau: let $i$ be the least
    index such that $c_i=c_{i+1}$.  Define a configuration $d$ as
    follows: $d_i=c_i-1$, $d_{i+1}=c_{i+1}+1$ and $d_j=c_j$ for
    $j\in[1,k]\setminus\set{i,i+1}$. Clearly, $c=\Vm_i(d)$ and
    $L(d)=L(c)$, $R(d)=R(c)$ is a crazed LR-decomposition by
    hypothesis. Again, if $i=1$ and $c_1 = 1$, $d$ has to be shifted
    by $1$. Let $m=\max L(c)$, then $L(d)=L(c)\setminus\set{m}$,
    $R(d)=(R(c)\cup\set{m})\setminus\set{k}$ is a crazed
    LR-decomposition of $d$.
  \item $L(c)$ contains no plateau: define a configuration $d$ such
    that $d_1 = c_1-1$, $d_2=c_2+1$ and $d_i=c_{i}$ for $i\in[3,k]$.
    Clearly, $c=\Vm_2(d)$, and $L(d)=L(c)$, $R(d)=R(c)$ is a crazed
    LR-decomposition of $d$. Once more, if $c_1 = 1$, let $m=\max
    L(c)$, $L(d)=L(c)\setminus\set{m}$ and
    $R(d)=(R(c)\cup\set{m})\setminus\set{k}$ for the same result.
  \end{itemize}
\end{proof}

The next proposition proves that having a crazed LR-decomposition is
sufficient to belong to an orbit graph.

\begin{proposition}\label{prop:crazT4.ogn}
  If a configuration $c$ admits a crazed LR-decomposition, then there
  is a $n \in \N$ such that $c \in \og{(n)}$.
\end{proposition}
\begin{proof}
  If $c=(n)$ for some $n\in\N$ then we are done.  Now, assume that
  $c=(c_1, \ldots, c_k)$ is such that $c\ne(n)$ with $n=\sum_{i=1}^k
  c_i$. Using Lemma~\ref{lem:cracra-1}, build a sequence of
  configurations $d^0, d^1,\ldots, d^h, \ldots$ such that $d^0=c$,
  $d^h\in\fb{d^{h+1}}$ for $h>0$ and $d^h$ admits a crazed
  LR-decomposition.  Remark that this sequence must be finite. Indeed,
  for all $h \in \N$, by Lemma~\ref{lem:Ef.inf.E}, $E(d^{h+1})>E(d^h)$
  and, by Lemma~\ref{lem:pre.nrg}, $E((n))\geq E(d^h)$ if $d^h\ne
  (n)$. Therefore there is $l \in \N$ such that $d^l = (n)$, hence
  there is a path in $\og{(n)}$ from $d^l = (n)$ to $d^0 = c$.
\end{proof}

Because of Proposition~\ref{prop:crazedLTR}, any fixed point $\Pi$ of
$\og{(n)}$ has very precise characteristics. It admits a crazed
LR-decomposition $L(\Pi), R(\Pi)$, and it has no cliffs. Therefore,
both $L(\Pi)$ and $R(\Pi)$ have at most $1$ plateau since they are
crazed.  Moreover, there may be another plateau at the junction
between $L(\Pi)$ and $R(\Pi)$, \ie at most 3 plateaus in $\Pi$.

\begin{figure}[!ht]
  \begin{center}
    \subfigure[Case $|T(\Pi)|=1$.]{
      \includegraphics[width=0.45\textwidth]{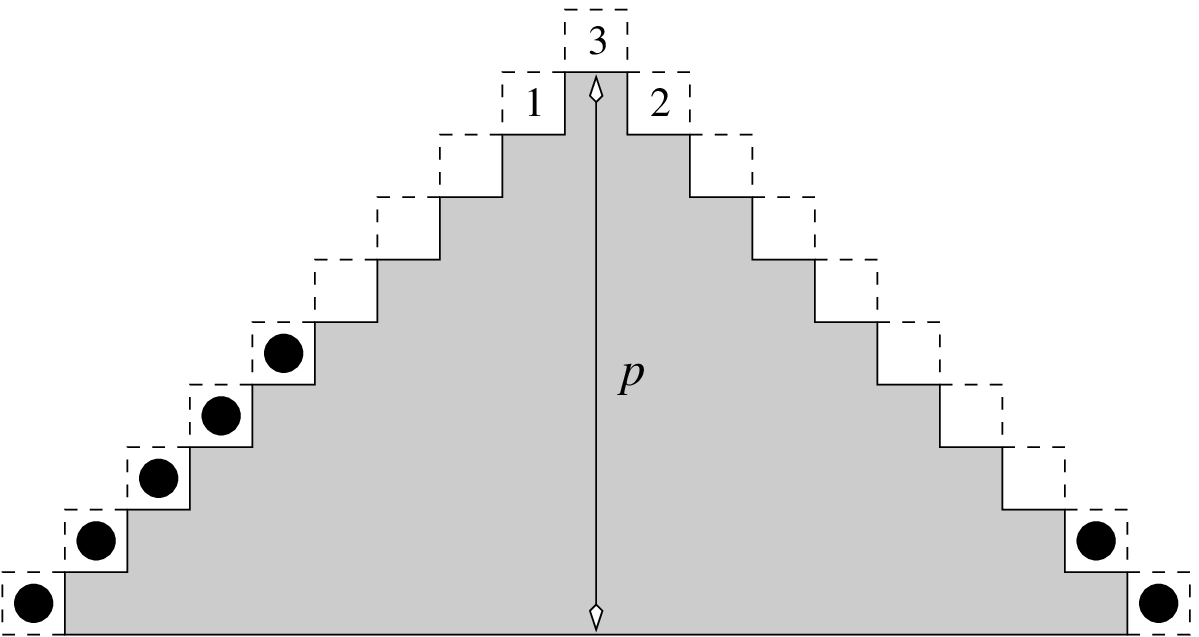}
      \label{fig:l1g1}
    }
    \subfigure[Case $|T(\Pi)|\geq 2$.]{
      \includegraphics[width=0.45\textwidth]{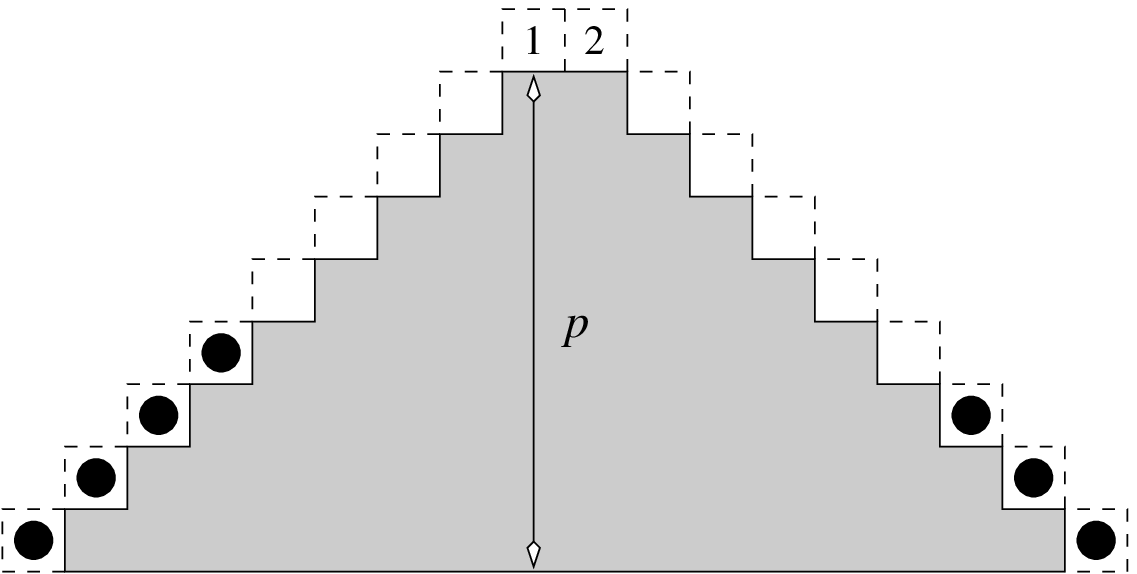}
      \label{fig:l2g2}
    }
    \caption{Structure of the fixed points.}
    \label{fig:lg}
  \end{center}
\end{figure}

The structure of the fixed points is described on
Figures~\ref{fig:lg}. Figure~\ref{fig:l1g1} represent the fixed points
$\Pi$ such that $|T(\Pi)| = 1$, Figure~\ref{fig:l2g2} is for the fixed
points $\Pi$ such that $|T(\Pi)| \geq 2$.

\subsection{A kind of magic}

From Corollary~\ref{cor:fixpstress}, we know that for any $n \in \N$, the
configuration $(n)$ leads to at least one fixed point. In this section
we compute precisely the number of fixed points of SSPM with initial
condition $(n)$.
\smallskip

In order to understand how a fixed point can be obtained, we try to
give a visual construction. Consider
Figures~\ref{fig:lg}. The $n$ grains of the fixed point must be arranged 
in the grayed part and can partially occupy the dashed frame with the 
supplementary constraint that grains in the dashed part must be as 
much clustered to the ground as possible. 
Boxes labeled $1$, $2$ and $3$ in Figures~\ref{fig:l1g1} and~\ref{fig:l2g2}
cannot be filled (for more details see the proof of Lemma~\ref{lem:G}). 
Remark that if $p$ is the height of the grayed
part, then this area contains $p^2$ grains in Figure~\ref{fig:l1g1},
and $p^2+p$ grains in Figure~\ref{fig:l2g2}. Appendix~\ref{app:fp}
shows all the possible fixed points reachable from the initial
condition $(n)$ for $n \in [1,32]$.

Lemma~\ref{lem:G} will be the main tool that we use to count the
number of fixed points, it proves that the ``shapes'' outlined in
Figure~\ref{fig:l2g2} describe
exactly all the possible fixed points. Let $g_1(n)$ be the numbers of
fixed points $\Pi$ such that $|T(\Pi)| = 1$, and $g_2(n)$ the numbers
of fixed points $\Pi$ such that $|T(\Pi)| \geq 2$.

\begin{lemma} \label{lem:G} For any $n\in\N$, consider SSPM with
  initial condition $(n)$.  The number of fixed points of $\og{(n)}$
  is given by $G(n)=g_1(n)+g_2(n)$.
\end{lemma}
\begin{proof}
  Let $n \in \N$, and $\Pi$ be a fixed point of $(n)$. If $|T(\Pi)| =
  1$, clearly $\Pi$ can be constructed as shown in Figure~\ref{fig:l1g1}
  (at most one plateau on the left and one on the right). Moreover, it
  cannot be constructed from Figure~\ref{fig:l2g2} since the boxes at
  the top (labelled $1$ and $2$) are left empty.

  If $|T(\Pi)| \geq 2$, the fixed point is not represented on
  Figure~\ref{fig:l1g1}, since the boxes labelled $1$ and $2$ cannot
  be filled. To show that it is constructible from
  Figure~\ref{fig:l2g2}, let $L(\Pi)=[1,t]$ and $R(\Pi)=[t+1, k]$ be a
  crazed decomposition of $\Pi=(\Pi_1, \ldots, \Pi_k)$. If $\Pi_t =
  \Pi_{t+1}$, cut Figure~\ref{fig:l2g2} in two parts at the middle of
  the configuration. It is clear that $L(\Pi)$ fits in the left part
  (at most one plateau), and $R(\Pi)$ fits in the right part for the
  same reason. If $\Pi_t = \Pi_{t+1}+1$, $T(\Pi) \subset L(\Pi)$. Cut
  Figure~\ref{fig:l2g2} in two parts, at the right of the two grains
  on top of the grayed pile, $L(\Pi)$ and $R(\Pi)$ fit. The
  symmetrical case is similar.

  Conversely, all the configurations with $n$ grains constructible
  from Figures~\ref{fig:lg} are clearly fixed points, are
  LR-decomposable, and hence are fixed points of $(n)$
  (Proposition~\ref{prop:crazT4.ogn}). Therefore, the total number of
  fixed points is the number of configurations constructible from
  Figure~\ref{fig:l1g1} plus the number of configurations
  constructible from Figure~\ref{fig:l2g2}, \ie $G(n) = g_1(n) +
  g_2(n)$.
\end{proof}

The two following lemmas give the exact expression of $g_1(n)$ and $g_2(n)$.

\begin{lemma}\label{lem:g.un}
   For any $n\in\N$, consider SSPM with initial condition $(n)$.
   The number of fixed points of $\og{(n)}$
   with top of length $1$ is given by
   \[
      g_1(n)=
      \begin{cases}
         n-p^2+1 & \text{if}\;n-p^2\leq p-1\enspace,\\
         2p-n+p^2-1 & \text{if}\;p\leq n-p^2\leq 2p-1\enspace,\\
         0 & \text{otherwise,}
      \end{cases}
   \]
   where $p$ is the unique integer such that $p^2\leq n < (p+1)^2$.
\end{lemma}
\begin{proof}
  For $n\in\N$, consider a fixed point $\Pi \in \og{(n)}$. Since by
  hypothesis $|T(\Pi)=1|$, the overall structure is illustrated in
  Figure~\ref {fig:l1g1}.  As $n$ is fixed, we can determine $p$: it
  is the unique integer satisfying $p^2 \leq n < (p+1)^2$, \ie
  $p=\lfloor \sqrt{n} \rfloor$. Now, let $u=n-p^2$ be the number of
  grains left after having arranged the grayed zone.  Distributing
  these $u$ grains consecutively and in all possible manners on the
  borders of the grayed zone starting from bottom to top gives all
  possible fixed points (lemma~\ref{lem:G}).  To be more precise we
  must distinguish three cases:
  \begin{itemize}
  \item $0 \leq u < p$: we put all $u$ grains in the free boxes on the
    left, from bottom to top; this gives a fixed point. Then we put
    only $u-1$ grains on the left and $1$ in the free box at the
    bottom on the right; this gives another fixed point. This process
    is iterated until there are $0$ gains on the left and $u$ on the
    right. It is clear that this procedure gives $u+1$ fixed points.
  \item $p \leq u \leq 2p-2$: we start by putting $p-1$ grains in the
    free boxes on the left, and the remaining $u-p+1$ grains in the
    free boxes on the right, starting from bottom to top; this gives a
    fixed point.  Then we put only $p-2$ grains on the left and the
    $u-p+2$ remaining grains in the free boxes on the right,
    proceeding from bottom to top; this gives another fixed point.
    Then, we can put $p-3$ grains on the left and so on until there
    are $p-1$ grains on the right. It is clear that $(p-1)-(u-p+1) + 1
    = 2p-u-1$ distinct fixed points can be generated in this manner.
  \item $u = 2p-1$ or $u=2p$; then there are necessarily $p$ grains on
    the left or on the right, hence there is one grain in box 1 or 2
    (Figure~\ref{fig:l1g1}). This should not be allowed, as it would
    mean that $|T(\Pi)| > 1$, which is not the case. In this last
    case, there are $0$ fixed points.
  \end{itemize}
  Finally, remark that box number $3$ is not taken into account
  either, since it would mean that all dashed boxes are filled and
  therefore we would have chosen $p+1$ as height of the pile instead
  of $p$.
\end{proof}

\begin{lemma}\label{lem:g.deux}
   For any $n\in\N$, consider SSPM with initial condition $(n)$.
   The number of fixed points of $\og{(n)}$
   with top of length bigger than $1$ is given by
   \[
      g_2(n)=
      \begin{cases}
         n-p^2-p+1 & \text{if}\;n-p^2-p\leq p-1\enspace,\\
         p & \text{if}\;n-p^2-p = p\enspace,\\
         3p-n+p^2+1 & \text{if}\;p+1\leq n-p^2-p\leq 2p+1\enspace,\\
      \end{cases}
   \]
   where $p$ is the unique integer such that $p^2+p\leq n < (p+1)^2 + (p+1)$.
\end{lemma}
\begin{proof}[Proof of~\ref{lem:g.deux}]
  The proof is similar to the one of Lemma~\ref{lem:g.un}.  For
  $n\in\N$, consider a fixed point $c\in\og{(n)}$. By hypothesis
  $|T(c)| \geq 2$, therefore the overall structure is the one
  illustrated in Figure~\ref {fig:l2g2}. Since $n$ is fixed, we can
  determine $p$: it is the unique integer satisfying $p^2+p\leq n <
  (p+1)^2+(p+1)$.  Let $v=n-p^2-p$ be the number of grains left after
  having arranged the grayed zone. Distributing the $v$ grains in all
  possible ways gives the number of fixed points $\Pi$ with $|T(\Pi)|
  \geq 2$ (Lemma~\ref{lem:G}). Again, we must distinguish three cases:
  \begin{itemize}
  \item $0 \leq v < p$: we put all $v$ grains in the free boxes on the
    left from bottom to top; this gives a fixed point. Then we put
    only $v-1$ grains on the left and $1$ in the free box on the right
    at the bottom; this gives another fixed point. This can be
    iterated until there are 0 grains on the left and $v$ on the
    right. It is clear that this procedure gives $v+1$ fixed points.
  \item $v = p$: again, we can put $p$ grains on the left, $0$ on the
    right and so on until there are $0$ on the left, $p$ on the right.
    Therefore there should be $p+1$ fixed points, but in fact the
    first one and the last one are exactly the same: top of length 3,
    and no plateaus. What happens is that the reference column is not
    at the same position, it is shifted by one, which does not matter.
    This is the only case of duplicated fixed point.
  \item $p < v \leq 2p+1$: we start by putting $p$ grains in the free
    boxes on the left, and the remaining $v-p$ grains in the free
    boxes on the right, starting from bottom to top; this gives a
    fixed point.  Then we put only $p-1$ grains on the left and the
    $v-p+1$ remaining grains on the free boxes to the right, proceeding
    from bottom to top; this gives another fixed point.  We proceed
    until there are $p$ grains on the right, it is clear that
    $p-(v-p)+1 = 2p-v+1$ distinct fixed points can be generated in
    this manner.
  \end{itemize}
  Remark that boxes at the top of the dashed pile must not be taken
  into account in the computation of the number of fixed points for it
  would mean that all dashed boxes are filled and therefore we would
  have chosen $p+1$ as height of the pile and not $p$. Moreover, if
  only one of them is filled, we generate a fixed point already
  counted in Lemma~\ref{lem:g.un} (top of length $1$).
\end{proof}

The following proposition gives a closed formula for the number of
fixed points in the orbit of initial condition $(n)$. The formula is
somewhat ``magical'' since it is very simple but we have 
neither  practical  nor visual explanation for it.

\begin{proposition} \label{prop:racine}
  For any $n\in\N$, consider SSPM with initial condition $(n)$.  The
  number of fixed points of $\og{(n)}$ is given by $G(n)=\lfloor
  \sqrt{n} \rfloor$.
\end{proposition}
\begin{proof}
  Let $n \in \N$ and $p=\lfloor \sqrt{n} \rfloor$ the only integer
  such that $p^2 \leq n < (p+1)^2$, there are three cases which
  correspond to the cases of Lemmas~\ref{lem:g.un}
  and~\ref{lem:g.deux}.
  \begin{itemize}
  \item If $p^2 \leq n \leq p^2+p-1$; then $(p-1)^2 + (p-1) \leq n <
    p^2+p$. This is case $1$ for Lemma~\ref{lem:g.un} and case $3$ for
    Lemma~\ref{lem:g.deux}, hence from Lemma~\ref{lem:G}, $G(n) =
    [n-p^2 + 1] + [3(p-1) - n + (p-1)^2 + 1] = p = \lfloor
    \sqrt{n} \rfloor$.
  \item If $p^2+p \leq n \leq p^2+2p-1$; we are in case $2$ of
    Lemma~\ref{lem:g.un} and case $1$ of Lemma~\ref{lem:g.deux}, hence
    from Lemma~\ref{lem:G}, $G(n) = [2p-n+p^2-1] + [n-p^2-p+1] = p =
    \lfloor \sqrt{n} \rfloor$.
  \item If $n = p^2+2p$, this is case $3$ for Lemma~\ref{lem:g.un} and
    case $2$ for Lemma~\ref{lem:g.deux}. From Lemma~\ref{lem:G},
    we find $G(n) = 0 + p = \lfloor \sqrt{n} \rfloor$.
  \end{itemize}
\end{proof}

Remark that it would also be possible to give the exact expression of
each of these fixed points, but it would be complex and of no interest
here. To have an idea of what they look like, please refer to  
Appendix~\ref{app:fp}.
\smallskip

Finally, remark that we did not take into account the initial position
of the columns. For the same fixed point, there may exist different
fixed points which have the same shape, but at different indices. In
this paper we do not consider this fact, we only take into account the
general shape of the configurations.

\section{Conclusions and future work}

In this paper we have introduced SSPM: a symmetric version of 
the well-known SPM model. We have proved that SSPM has
fixed point dynamics and exhibited the precise structure of
the fixed points which are in the orbit of initial condition $(n)$.
Moreover, we showed a simple closed formula for counting the number
of distinct (\ie having different shape) fixed points. Remark that this result
is surprising since the combinatorial complexity of the orbit graphs becomes
higher and higher when the number $n$ of grains grows. This complexity
contrasts with the simplicity of the formula for the number of fixed points:
$\left\lfloor\sqrt{n}\right\rfloor$.
Moreover, this formula is to some extent fascinating: although it is very simple,
we have neither a practical nor a visual explanation for it. 

This research can be continued along three main directions:
\begin{itemize}
  \item Corollary~\ref{cor:fixpstress}  says that, starting from any initial 
            configuration, SSPM has fixed point dynamics. Can we give a formula
            or at least tight bounds for the shortest path to a fixed point?
            For the longest? 
  \item Section~\ref{sec:og} gives a precise characterization of orbit graphs for
            initial conditions made of one single column. It would be interesting to
            extend this characterization to more general initial conditions or at least
            to find an alternative characterization.
  \item  The model we introduced is intrinsically sequential: only one grain moves  
            at each time step. It would be interesting to introduce a model similar to SSPM
            but with synchronous update. This would be even more realistic than SSPM for
            the simulation of natural phenomena.
\end{itemize}
\ignore{
In what concerns the third point,
it seems to be quite difficult to define a synchronous model. Indeed,  a grain may be able to move
on both sides at the same time. Moving 2 grains at the same time in both directions, or choosing a direction randomly, does not simplify the problem.
}
\bibliographystyle{plain}
\bibliography{sas}
\newpage
\appendix
\section{Fixed points of $(n)$ for $1 \leq n \leq 32$} \label{app:fp}
\begin{minipage}[b]{0.25\textwidth}\begin{center}
  \includegraphics[scale=0.43]{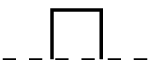} \\
  $n = 1$
\end{center}\end{minipage}\begin{minipage}[b]{0.25\textwidth}\begin{center}
  \includegraphics[scale=0.43]{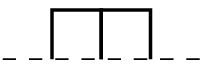} \\
  $n = 2$
\end{center}\end{minipage}\begin{minipage}[b]{0.25\textwidth}\begin{center}
  \includegraphics[scale=0.43]{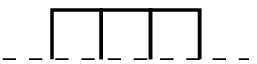} \\
  $n = 3$
\end{center}\end{minipage}\begin{minipage}[b]{0.25\textwidth}\begin{center}
  \includegraphics[scale=0.43]{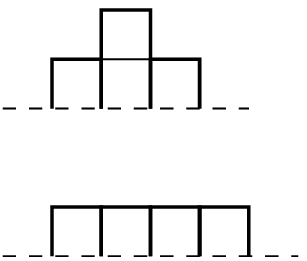} \\
  $n = 4$
\end{center}\end{minipage}\\[8ex]
\begin{minipage}[b]{0.25\textwidth}\begin{center}
  \includegraphics[scale=0.43]{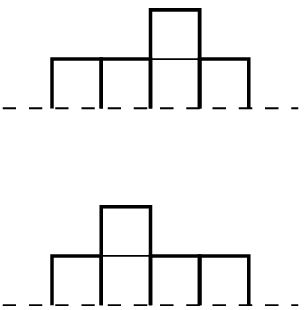} \\
  $n = 5$
\end{center}\end{minipage}\begin{minipage}[b]{0.25\textwidth}\begin{center}
  \includegraphics[scale=0.43]{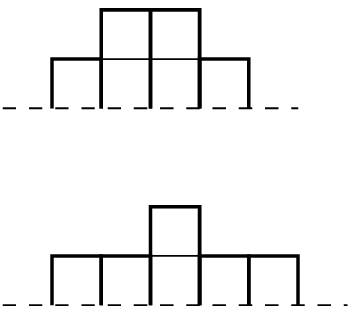} \\
  $n = 6$
\end{center}\end{minipage}\begin{minipage}[b]{0.25\textwidth}\begin{center}
  \includegraphics[scale=0.43]{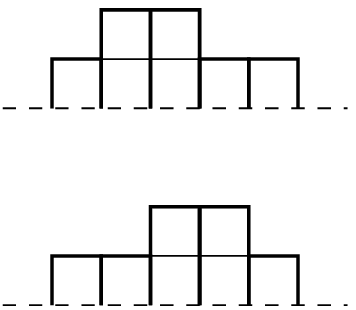} \\
  $n = 7$
\end{center}\end{minipage}\begin{minipage}[b]{0.25\textwidth}\begin{center}
  \includegraphics[scale=0.43]{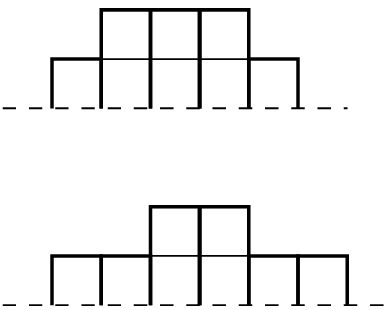} \\
  $n = 8$
\end{center}\end{minipage}\\[8ex]
\begin{minipage}[b]{0.25\textwidth}\begin{center}
  \includegraphics[scale=0.43]{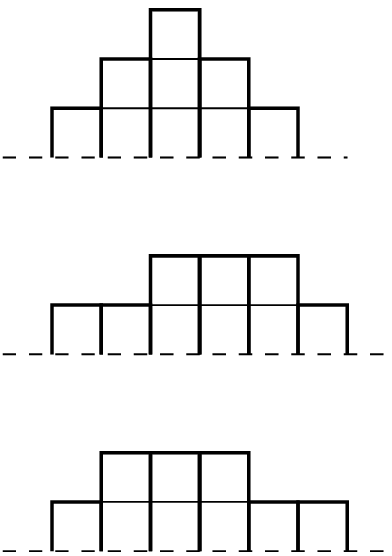} \\
  $n = 9$
\end{center}\end{minipage}\begin{minipage}[b]{0.25\textwidth}\begin{center}
  \includegraphics[scale=0.43]{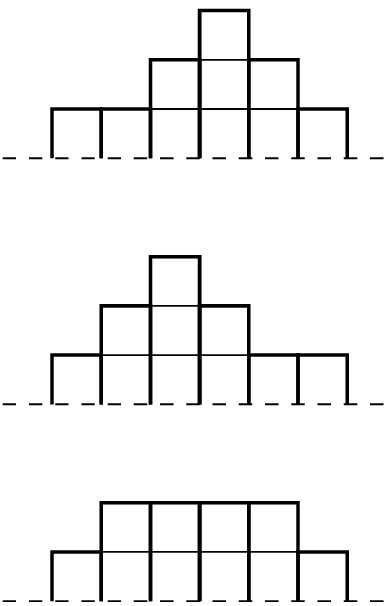} \\
  $n = 10$
\end{center}\end{minipage}\begin{minipage}[b]{0.25\textwidth}\begin{center}
  \includegraphics[scale=0.43]{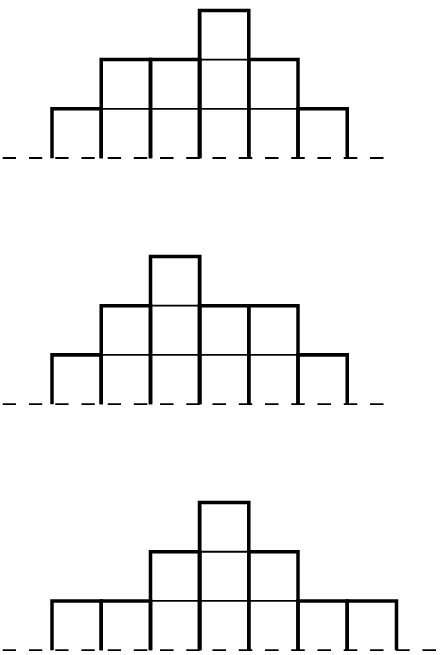} \\
  $n = 11$
\end{center}\end{minipage}\begin{minipage}[b]{0.25\textwidth}\begin{center}
  \includegraphics[scale=0.43]{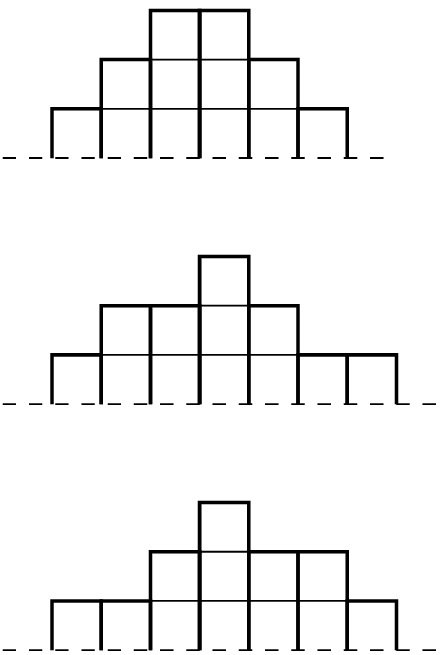} \\
  $n = 12$
\end{center}\end{minipage}\\[8ex]
\begin{minipage}[b]{0.25\textwidth}\begin{center}
  \includegraphics[scale=0.43]{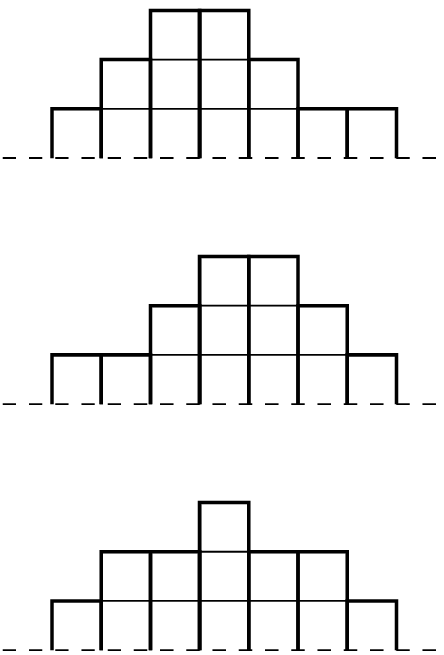} \\
  $n = 13$
\end{center}\end{minipage}\begin{minipage}[b]{0.25\textwidth}\begin{center}
  \includegraphics[scale=0.43]{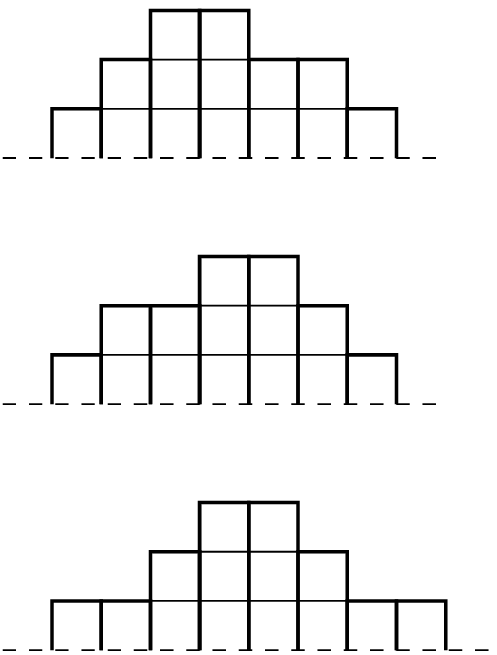} \\
  $n = 14$
\end{center}\end{minipage}\begin{minipage}[b]{0.25\textwidth}\begin{center}
  \includegraphics[scale=0.43]{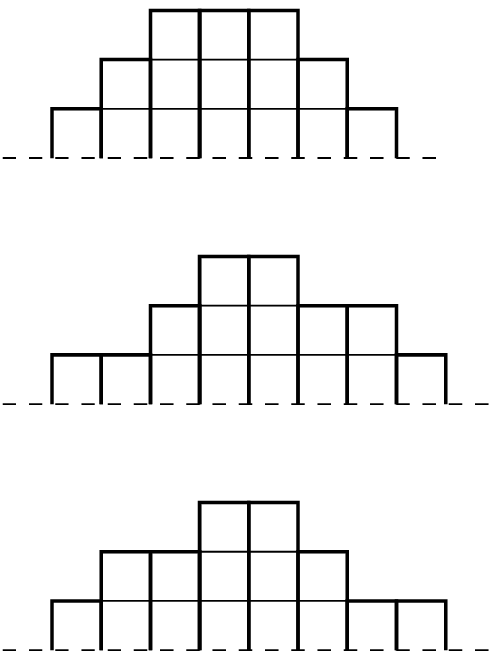} \\
  $n = 15$
\end{center}\end{minipage}\begin{minipage}[b]{0.25\textwidth}\begin{center}
  \includegraphics[scale=0.43]{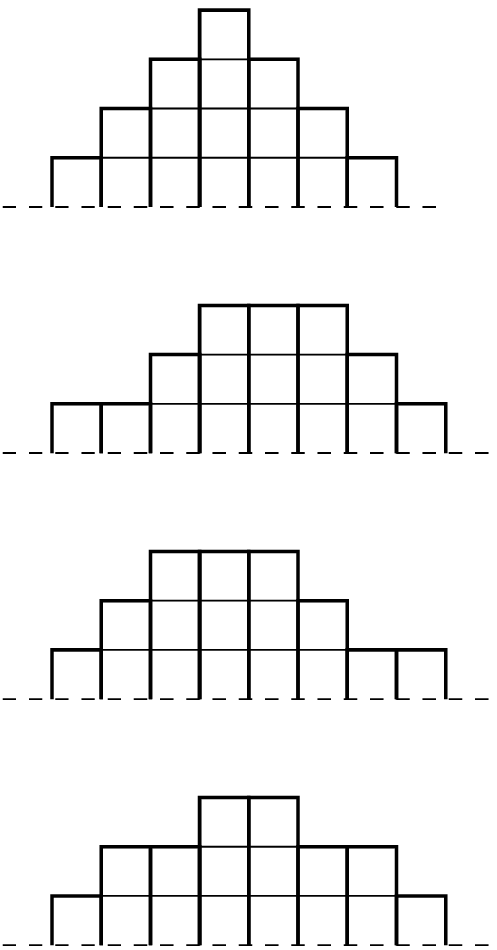} \\
  $n = 16$
\end{center}\end{minipage}\\[8ex]
\begin{minipage}[b]{0.25\textwidth}\begin{center}
  \includegraphics[scale=0.43]{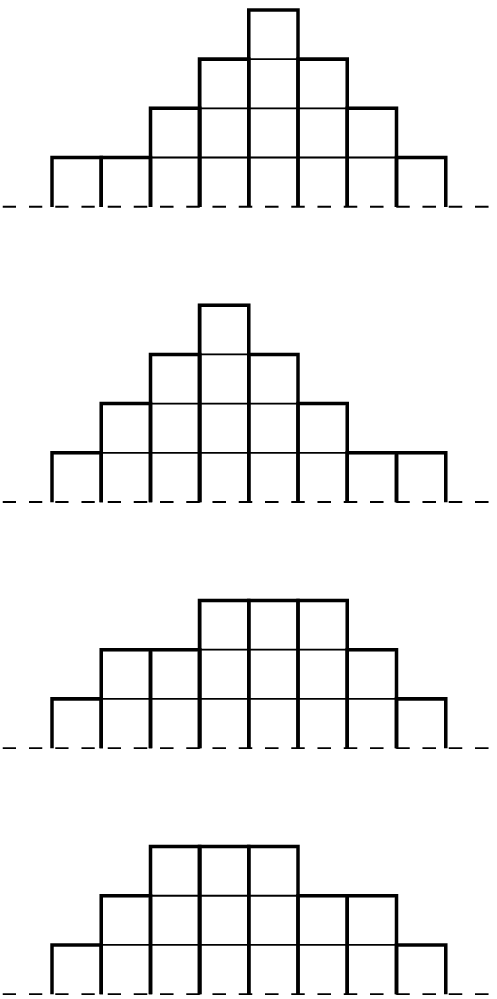} \\
  $n = 17$
\end{center}\end{minipage}\begin{minipage}[b]{0.25\textwidth}\begin{center}
  \includegraphics[scale=0.43]{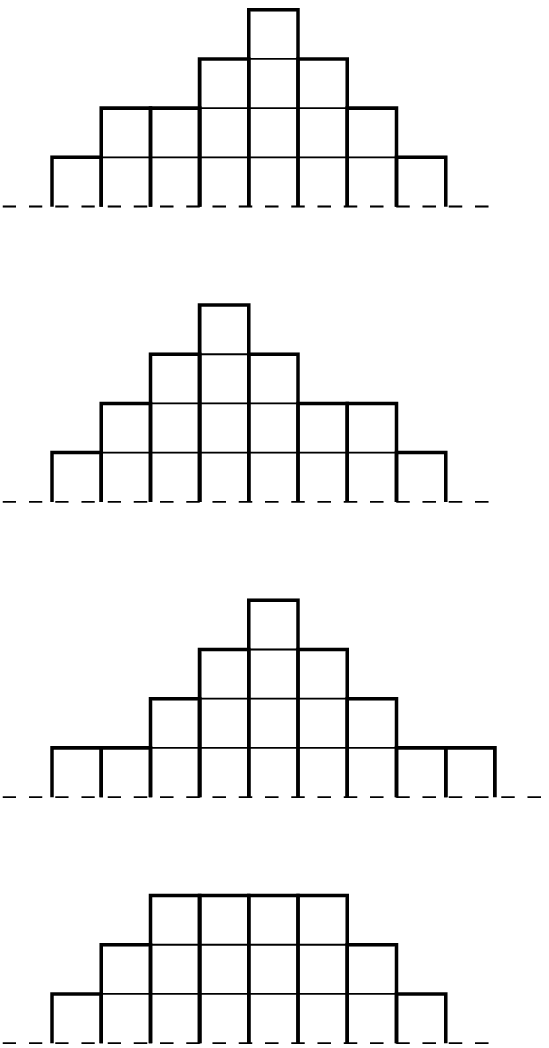} \\
  $n = 18$
\end{center}\end{minipage}\begin{minipage}[b]{0.25\textwidth}\begin{center}
  \includegraphics[scale=0.43]{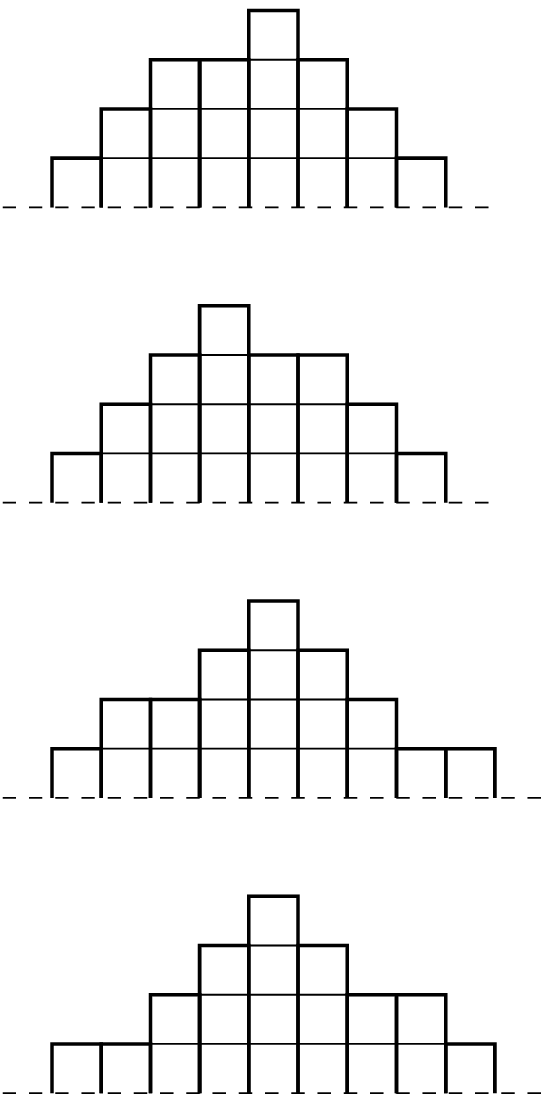} \\
  $n = 19$
\end{center}\end{minipage}\begin{minipage}[b]{0.25\textwidth}\begin{center}
  \includegraphics[scale=0.43]{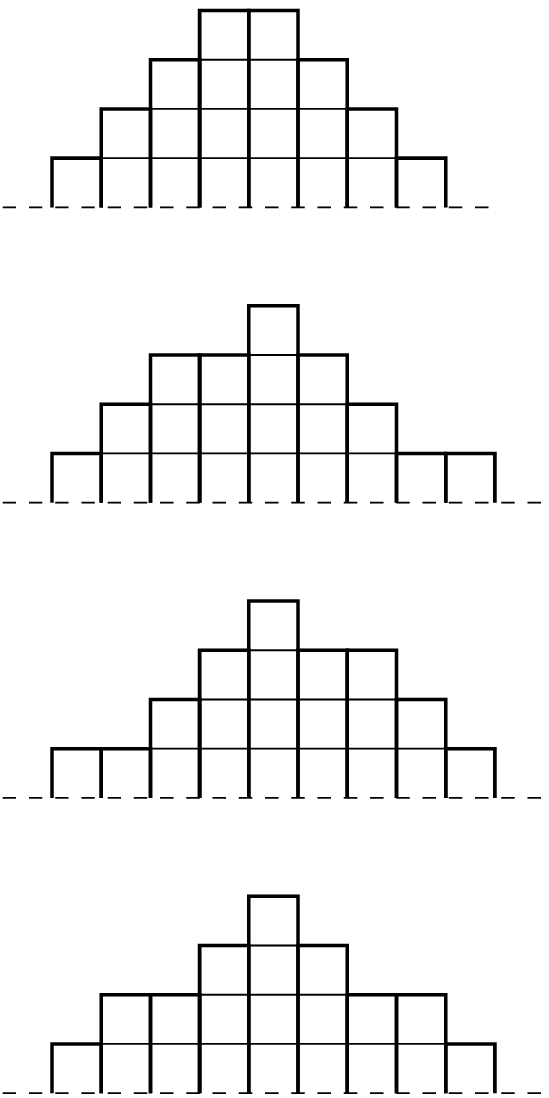} \\
  $n = 20$
\end{center}\end{minipage}\\[8ex]
\begin{minipage}[b]{0.25\textwidth}\begin{center}
  \includegraphics[scale=0.43]{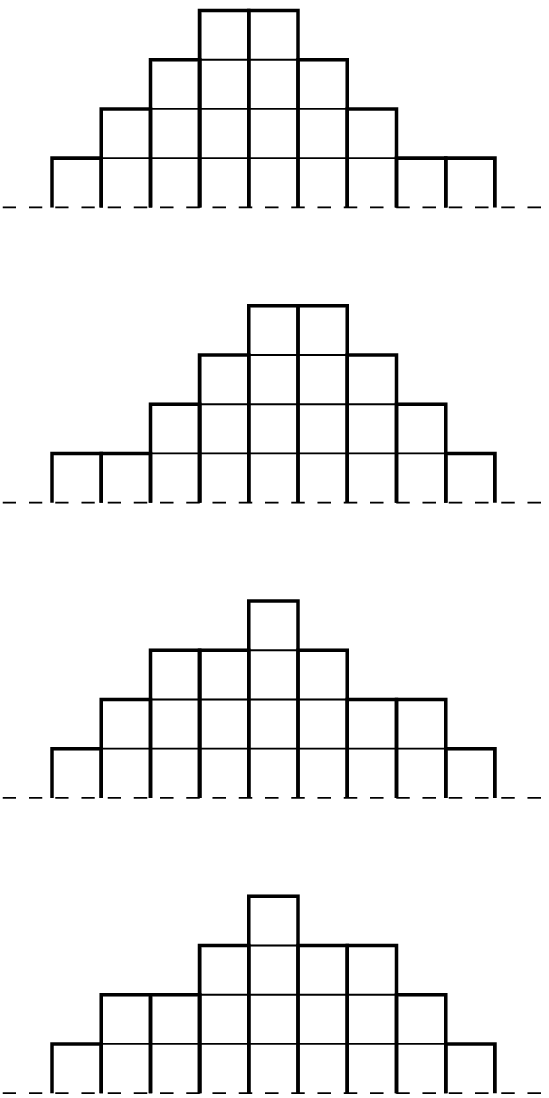} \\
  $n = 21$
\end{center}\end{minipage}\begin{minipage}[b]{0.25\textwidth}\begin{center}
  \includegraphics[scale=0.43]{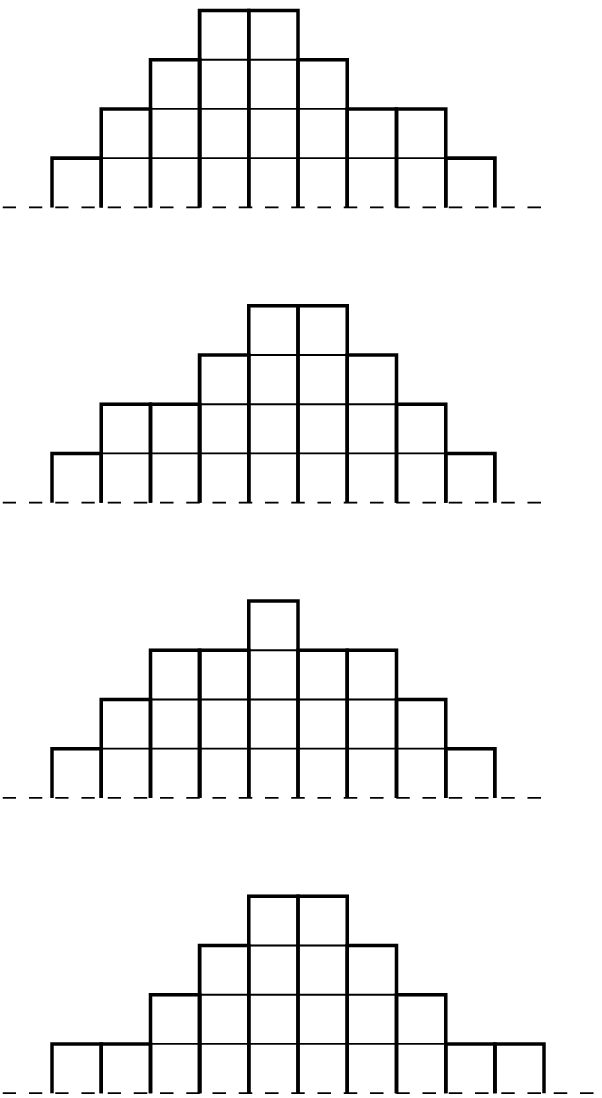} \\
  $n = 22$
\end{center}\end{minipage}\begin{minipage}[b]{0.25\textwidth}\begin{center}
  \includegraphics[scale=0.43]{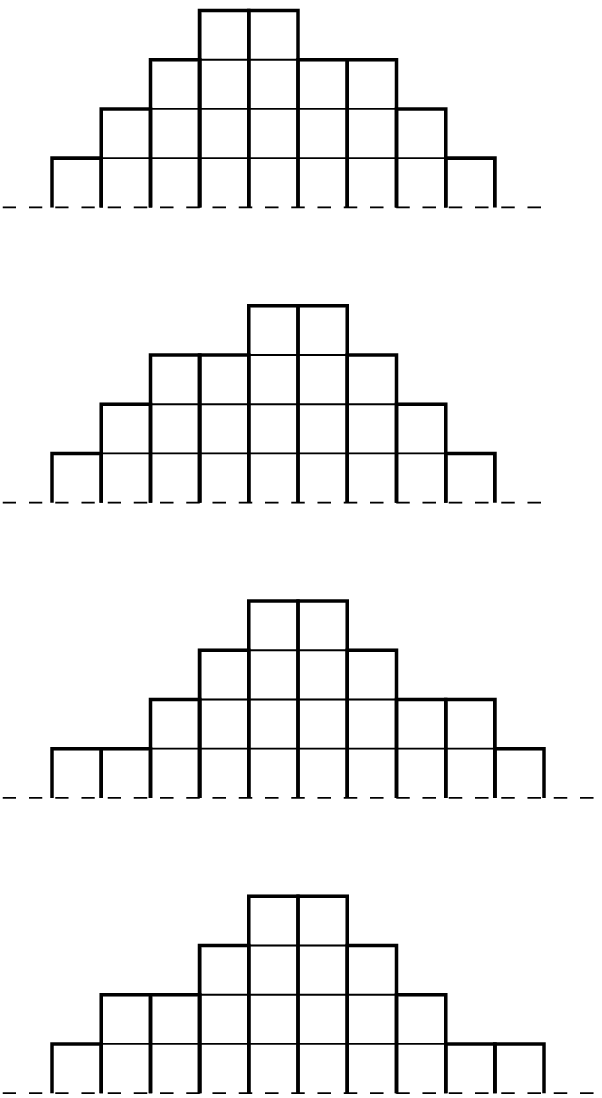} \\
  $n = 23$
\end{center}\end{minipage}\begin{minipage}[b]{0.25\textwidth}\begin{center}
  \includegraphics[scale=0.43]{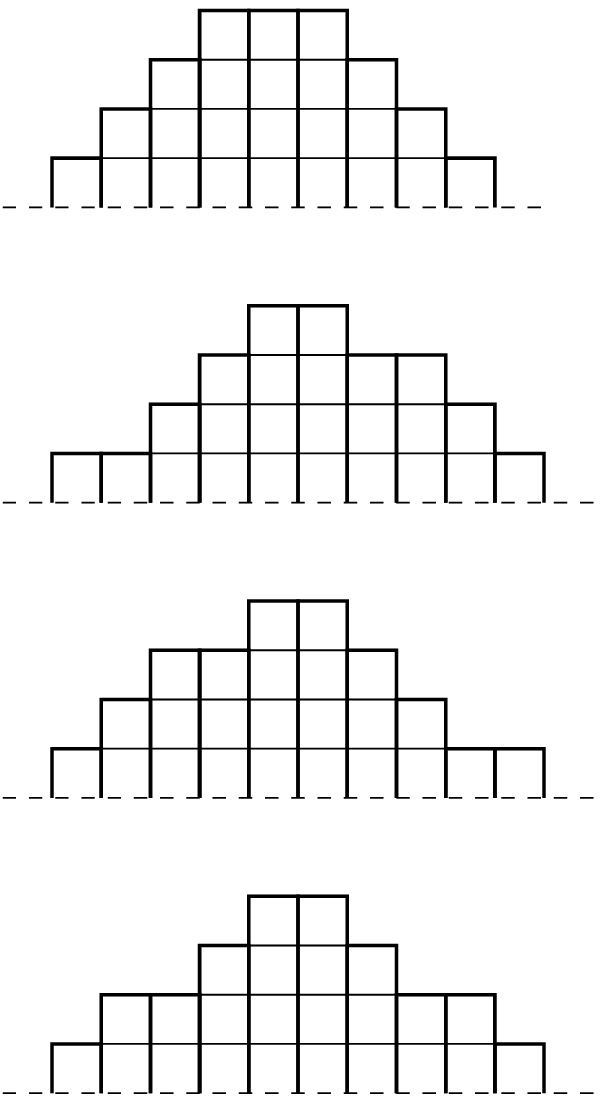} \\
  $n = 24$
\end{center}\end{minipage}\\[8ex]
\begin{minipage}[b]{0.25\textwidth}\begin{center}
  \includegraphics[scale=0.43]{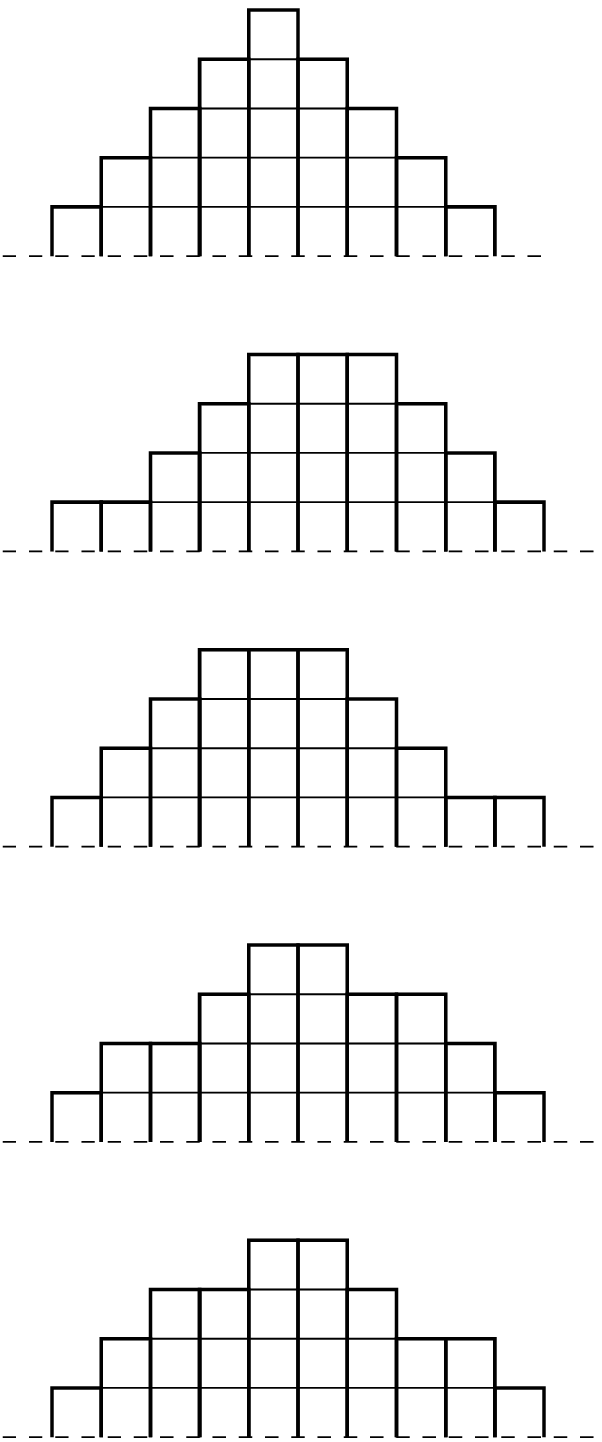} \\
  $n = 25$
\end{center}\end{minipage}\begin{minipage}[b]{0.25\textwidth}\begin{center}
  \includegraphics[scale=0.43]{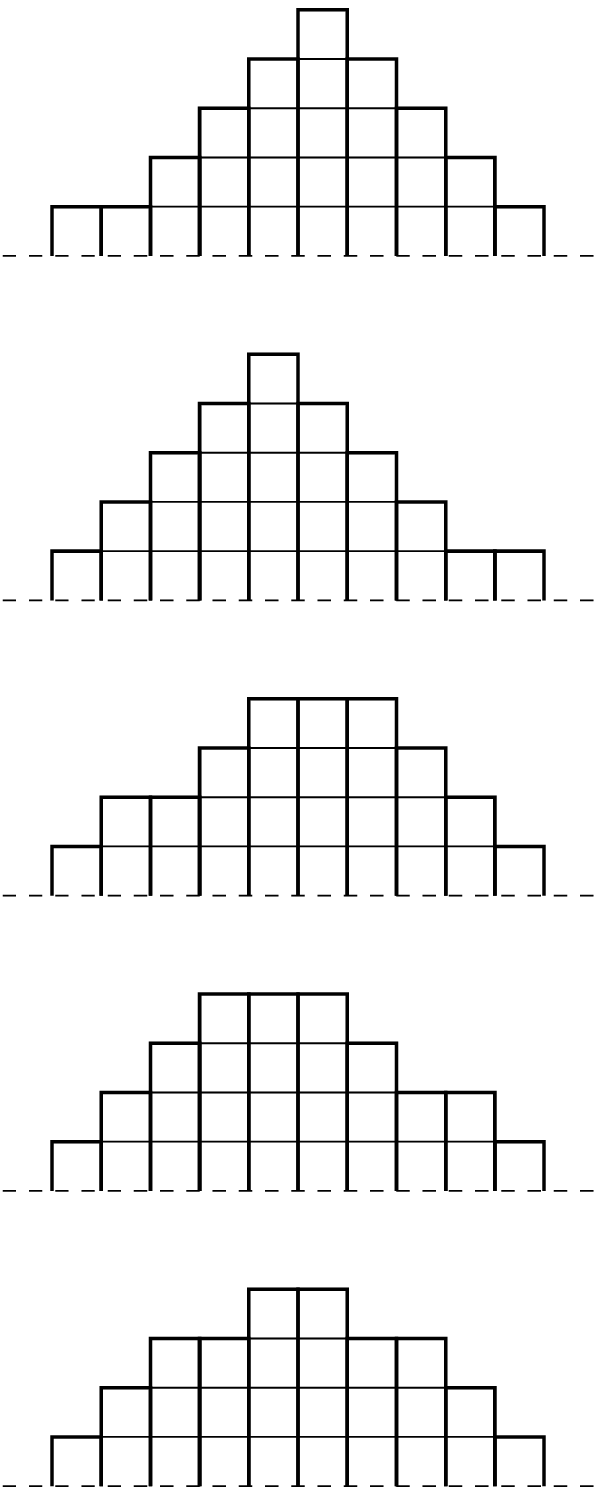} \\
  $n = 26$
\end{center}\end{minipage}\begin{minipage}[b]{0.25\textwidth}\begin{center}
  \includegraphics[scale=0.43]{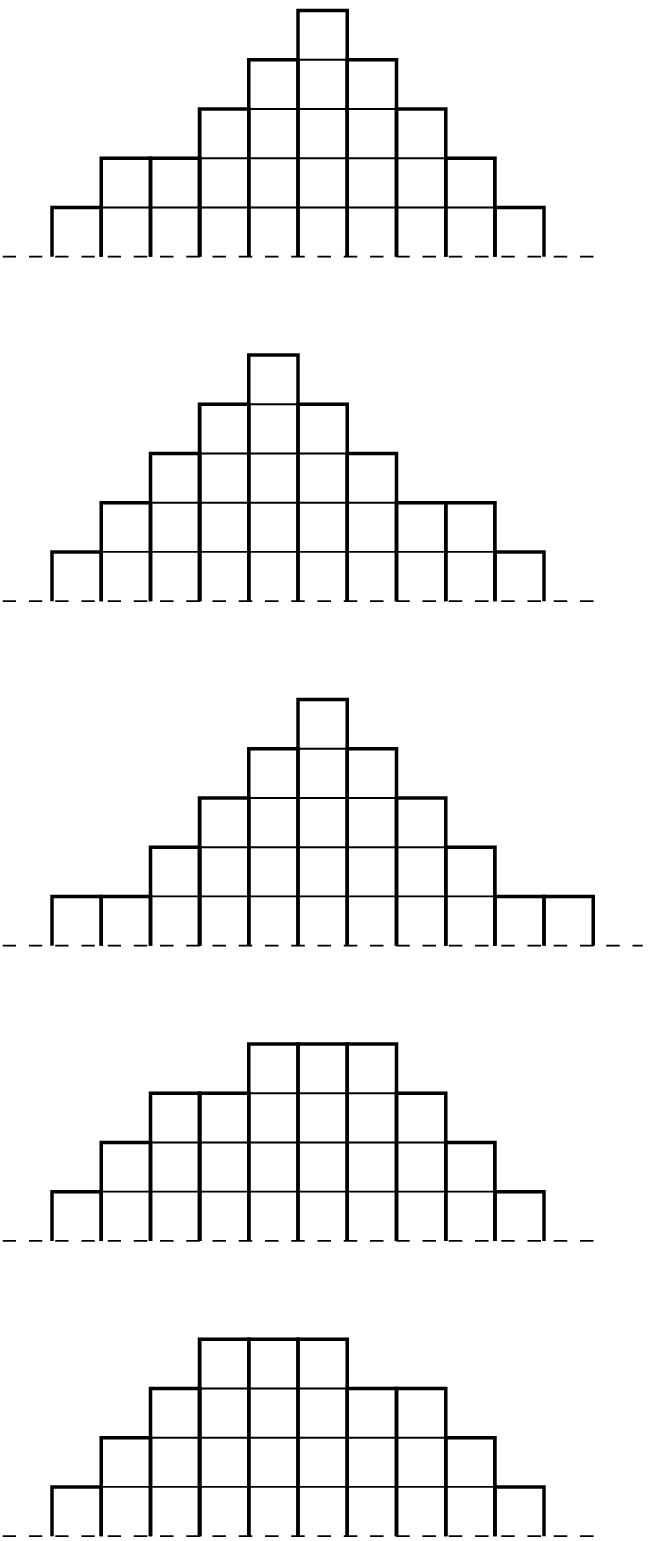} \\
  $n = 27$
\end{center}\end{minipage}\begin{minipage}[b]{0.25\textwidth}\begin{center}
  \includegraphics[scale=0.43]{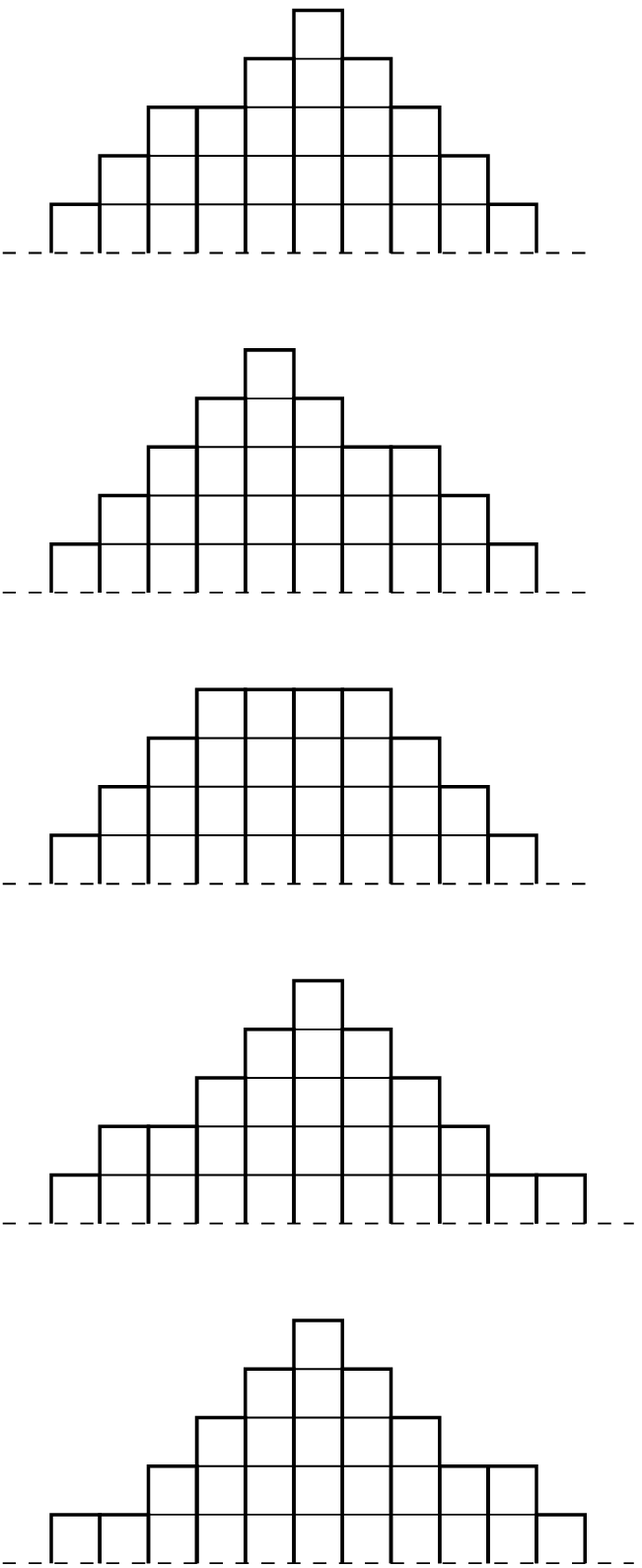} \\
  $n = 28$
\end{center}\end{minipage}\\[8ex]
\begin{minipage}[b]{0.25\textwidth}\begin{center}
  \includegraphics[scale=0.43]{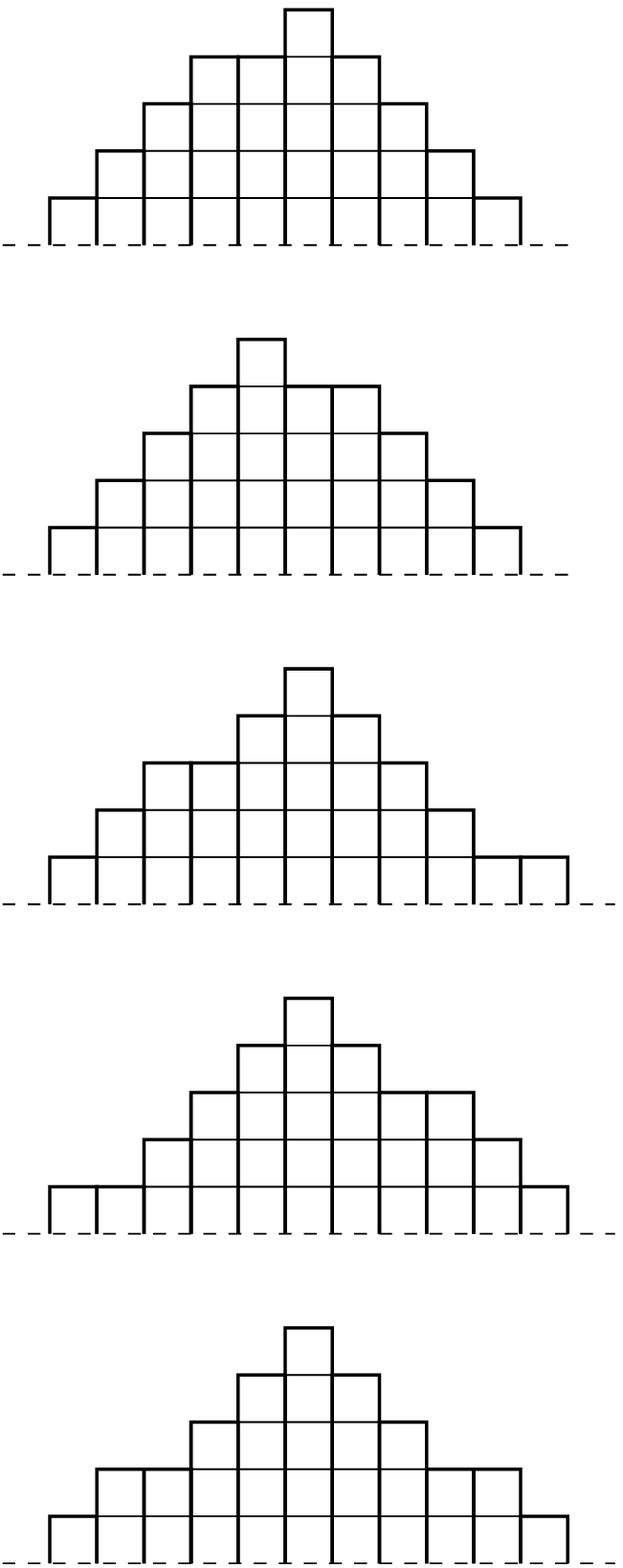} \\
  $n = 29$
\end{center}\end{minipage}\begin{minipage}[b]{0.25\textwidth}\begin{center}
  \includegraphics[scale=0.43]{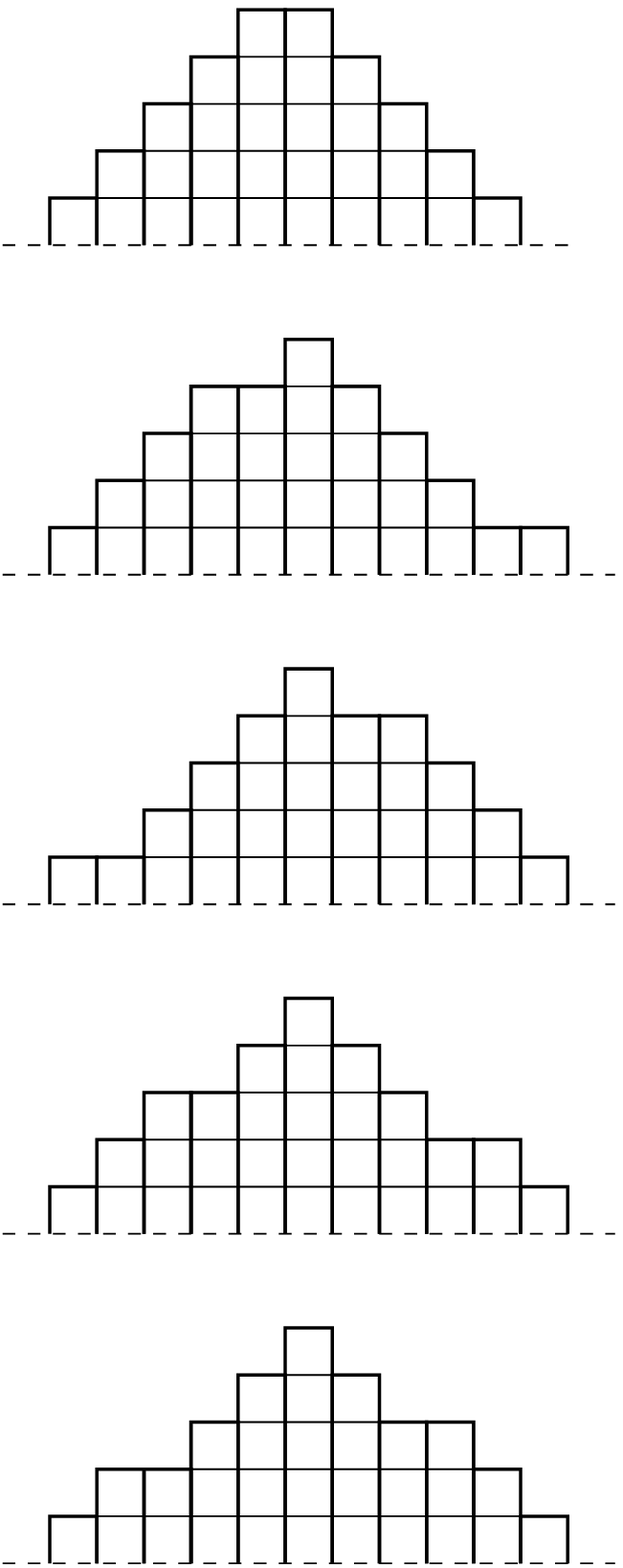} \\
  $n = 30$
\end{center}\end{minipage}\begin{minipage}[b]{0.25\textwidth}\begin{center}
  \includegraphics[scale=0.43]{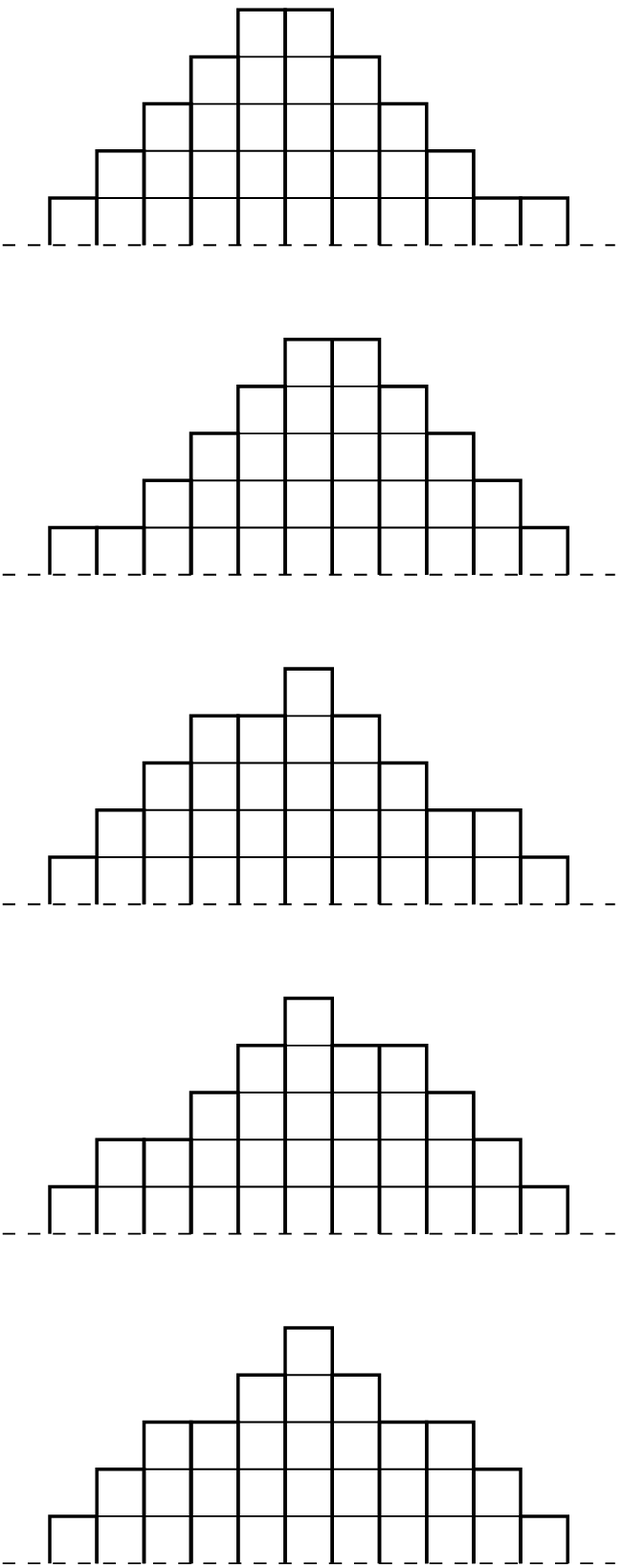} \\
  $n = 31$
\end{center}\end{minipage}\begin{minipage}[b]{0.25\textwidth}\begin{center}
  \includegraphics[scale=0.43]{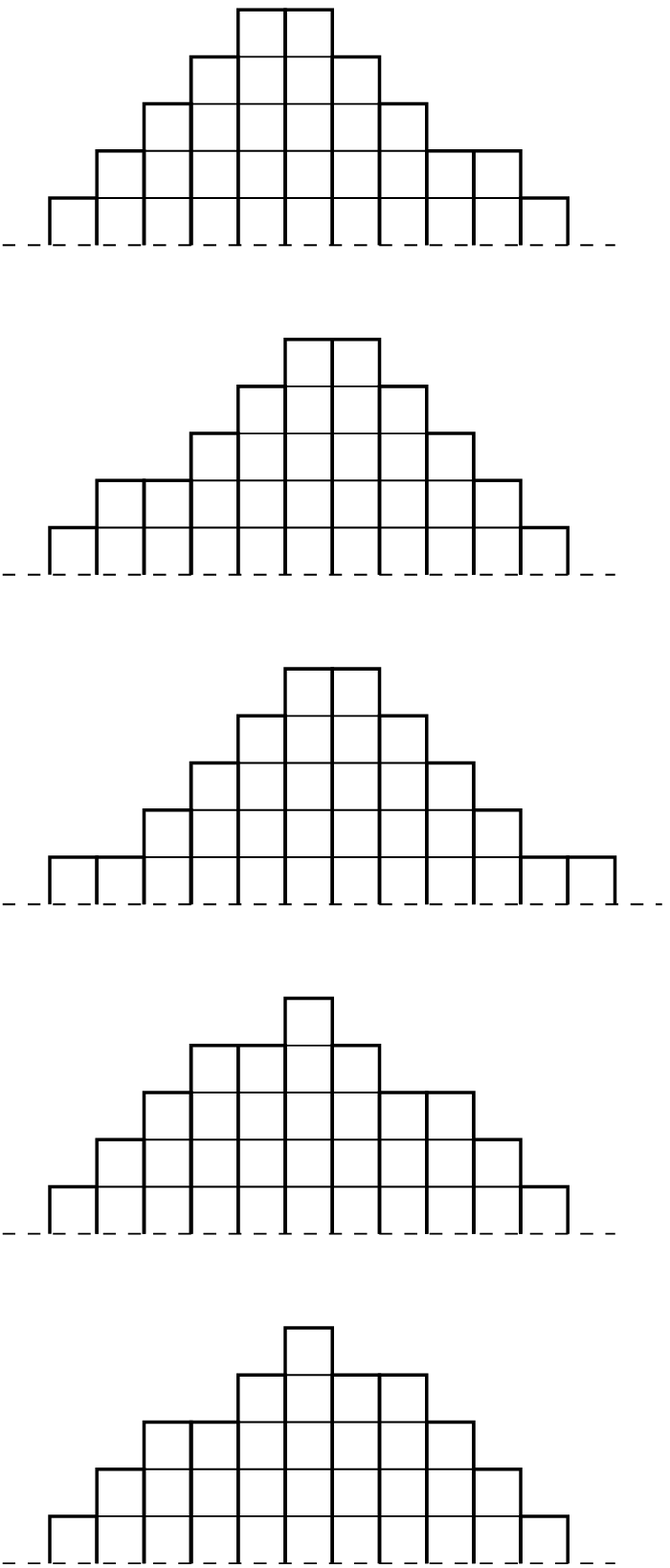} \\
  $n = 32$
\end{center}\end{minipage}

\end{document}